**Edible polysaccharides as stabilizers and carriers for the delivery of phenolic compounds and pigments in food formulations**

Liliane Siqueira de Oliveira[1,2] (https://orcid.org/0000-0002-2791-0325), Davi Vieira Teixeira da Silva[1,2] (https://orcid.org/0000-0003-2839-9984), Lucileno Rodrigues da Trindade[1,2] (https://orcid.org/0000-0002-1940-6698), Diego dos Santos Baião[1,2] (https://orcid.org/0000-0002-8278-7067), Cristine Couto de Almeida[1,2] (https://orcid.org/0000-0003-3529-0530), Vitor Francisco Ferreira[3] (https://orcid.org/0000-0002-2166-766X) and Vania Margaret Flosi Paschoalin[1,2,4*] (https://orcid.otg/0000-0001-6093-134X).

[1] Federal University of Rio de Janeiro (UFRJ), Department of Biochemistry, Chemistry Institute, Avenida Athos da Silveira Ramos 149 – sala 545 - Cidade Universitária - 21941-909 - Rio de Janeiro - RJ, Brazil. L.S.O. nutrililianesiqueira@gmail.com; D.V.T.S. davivieiraufrj@gmail.com; L.R.T. lucileno.trindade@gmail.com; D.S.B. diegobaiao20@hotmail.com; C.C.A. almeidacristine@hotmail.com; VFF. vitorferreira@id.uff.br; V.M.F.P. paschv@iq.ufrj.br.

[2] Graduate Studies in Food Sciences, Chemistry Institute, Federal University of Rio de Janeiro (UFRJ), Avenida Athos da Silveira Ramos, 149 – sala 545 – Cidade Universitária - 21941-909 - Rio de Janeiro – RJ, Brazil.

[3] Graduate Studies in Chemistry, Faculty of Pharmacy, Department of Pharmaceutical Technology, Federal Fluminense University (UFF), Rua Dr. Mario Vianna, 523 – Santa Rosa - 24210-141 - Niterói – RJ, Brazil.

[4] Graduate Studies in Chemistry, Chemistry Institute, Federal University of Rio de Janeiro (UFRJ), Avenida Athos da Silveira Ramos, 149 – sala 622 – Cidade Universitária - 21941-909 - Rio de Janeiro – RJ, Brazil.

* Correspondence: paschv@iq.ufrj.br; Phone number: +55(21)3938-7362; Fax number: +55(21)3938-7362

## ABSTRACT

Food polysaccharides have emerged as suitable carriers of active substances and as additives to food and nutraceutical formulations, showing potential to stabilize bioactive compounds during the storage of microencapsulate preparations, even in the gastrointestinal tract following the intake of bioactive compounds, thereby improving their bioaccessibility and bioavailability. This review provides a comprehensive overview of the main polysaccharides employed as wall materials, including starch, maltodextrin, alginate, pectin, inulin, chitosan, and gum arabic, and discusses how structural interactions and physicochemical properties can benefit the microencapsulation of polyphenols and pigments. The main findings and principles of the major encapsulation techniques, including spray drying, freeze drying, extrusion, emulsification, and coacervation, related to the production of microparticles, were briefly described. Polysaccharides can entrap hydrophilic and hydrophobic compounds by physical interactions, forming a barrier around the nucleus or binding to the bioactive compound. Intermolecular binding between polysaccharides in the wall matrix, polyphenols, and pigments in the nucleus can confer up to 90% of encapsulation efficiency, governed mainly by hydrogen bonds and electrostatic interactions. The mixture of wall polysaccharides in the microparticles synthesis favors the encapsulation solubility, storage stability, bioaccessibility, and bioactivity of the microencapsulate compounds. Clinical trials on the bioefficacy of polyphenols and pigments loaded in polysaccharide microparticles are scarce and require further evidence to reinforce the use of this technology.

## 1. INTRODUCTION

Microencapsulation refers to the use of different materials to entrap and stabilize solid, liquid, or gaseous substances within a nucleus, resulting in microscopic particles that protect, transport, and release active compounds in a controlled manner, depending on the properties of the coating agent (Ozkan et al., 2019). Microencapsulated compounds may exhibit enhanced nutraceutical and therapeutic effects due to improvements in the chemical stability against environmental conditions during storage, resulting in enhanced bioaccessibility and bioavailability after intake (Costa et al., 2021; Rezagholizade et al., 2024).

Microencapsulation can be employed for food and nutraceutical enrichment by embodying bioactive compounds, extending their half-life, avoiding sensory alterations, and preventing anti-nutritional interactions (Sarubbo et al., 2022; Rezagholizade et al., 2024).

Several edible polysaccharides derived from plants and agro-industrial residues are employed in food processing and derived formulations due to their low costs, biodegradability, and biocompatibility. In addition, these polysaccharides are components of the habitual diet and are generally recognized as safe (GRAS) by the United States Food and Drug Administration (FDA) (Benalaya et al., 2024). Starch, maltodextrin, pectin, gum arabic, xanthan gum, and sodium alginate, among others, can protect and facilitate bioactive compounds delivery, justifying their broad application as food-based carrier systems (Stevanovic & Filipović, 2024).

Bioactive compounds found in food matrices may perform metabolic or physiological roles that provide health benefits when ingested regularly, in addition to their role in mitigating the development of chronic diseases (Samborska et al., 2021). Intake of bioactive compounds can modulate lipid and glucose metabolism, improve endothelial function and blood pressure, and may promote oxidative homeostasis in the human body (Fraga et al., 2019). These benefits align with epidemiological studies, which corroborate the inverse association between regular fruit and plant intake and the reduced risk of cardiovascular diseases and cancer (Wang et al., 2021).

Naturally occurring antioxidants, as the polyphenols and pigments found in plant foods, are secondary metabolites that modulate multiple cellular processes capable of promoting human health (Duda-Chodak et al., 2015). Polyphenols comprise an extensive group of compounds, including phenolic acids, flavonoids, stilbenes, and lignans. In addition, several of these compounds are natural dyes in food, such as carotenoids, which impart red, yellow, and orange hues, yellow-colored flavonoids, green-colored chlorophylls, purple-colored anthocyanins, and red-violet and yellow betalains (Razem et al., 2022). Health benefits of polyphenols and pigments reside in their ability to neutralize reactive oxygen and nitrogen species and induce expression of genes encoding antioxidant and anti-inflammatory enzymes (Fraga et al., 2019; Rezagholizade et al., 2024). These compounds are marketed mainly, and their addition to foods and nutraceutical products is regulated by the FDA and the European Food Safety Authority (EFSA) agencies (De Almeida et al., 2024). To benefit from those natural compounds, it is necessary to overcome their poor water solubility, interactions with anti-nutritional dietary components, instability in the gastrointestinal tract following enzymatic and microbiological degradation, and instability in the bloodstream by plasma enzymes, which result in rapid clearance (Duda-Chodak et al., 2015).

Polysaccharide microparticles loaded with polyphenols and pigments have been developed to stabilize these compounds and are synthesized by different techniques, including spray drying, freeze-drying, extrusion, coacervation, and emulsification. Optimized conditions are required to achieve satisfactory microencapsulation and bioactive preservation, such as technique suitability, choice of wall polymer, and optimization of processing conditions (Ozkan et al., 2019). Interactions between functional groups of bioactive compounds and polysaccharides, such as hydrogen bonding and hydrophobic interactions with polyphenols and pigments, are critical factors that affect the physicochemical behavior of these encapsulated compounds and must be considered (Jakobek, 2014).

This narrative review describes the main findings concerning the synthesis of polysaccharide microparticles, evaluating the main edible polysaccharides for polyphenols and pigments delivery. We highlight the main findings considering the effect of microencapsulation on the physicochemical properties of polyphenols, such as solubility, storage stability, stability in gastrointestinal fluids, bioactivity, and encapsulation efficiency. In addition, the main techniques applied in the microparticles synthesis, the potential chemical interactions between the components of microparticles underlying the benefits of the microencapsulation, and perspectives on biological effects are addressed.

## 2. POLYSACCHARIDES AS VEHICLES FOR THE ENTRAPMENT OF POLYPHENOLS AND PIGMENTS FROM FOOD MATRICES

Several possibilities have been developed regarding the use of edible polysaccharides as coating materials, alone or as a mixture to build the wall matrix, demonstrating the high capacity to retain hydrophilic and hydrophobic bioactive compounds. This occurs mainly through physical interactions between molecules that form a polymeric barrier, as well as through interactions between the active nucleus and functional groups of microparticle components, increasing the structural stability and desirable physicochemical characteristics, and then improving functional properties for the enrichment of foods and nutraceuticals. The main findings on the physicochemical benefits conferred by polysaccharide microparticles coating phenolic compounds and pigments are summarized in the Table 1.

## 2.1 Starch

Starch is a polysaccharide composed of amylose and amylopectin polymers, which make up amorphous and crystalline starch structure areas, respectively (Figure 1). Amylose consists of a predominantly linear sequence of D-glucopyranose units linked by α-1,4-glycosidic bonds and amylopectin presents an essentially branched structure formed by α-1,6-glycosidic bonds (Buléon et al., 1998).

Starch is a widely employed polysaccharide in the food industry, mainly added to products as a thickening, emulsifying, gelling, and encapsulating agent, being obtained from cereals, roots, tubers, legumes, and certain unripe fruits, such as bananas and mangoes (Wang et al., 2022). In the food industry, starch is primarily extracted from corn and used as an additive to enhance the sensory attributes of processed foods, such as pastas, soups, sauces, and mayonnaises (Egharevba, 2019). However, starch exhibits low thermal stability, retrogradation resistance, and poor water solubility, motivating the use of modified starches to achieve the desired properties (Egharevba, 2019; Santana & Meireles, 2014). Starch is modified when native granules are subject to physical, chemical, or enzymatic processing. The chemical modification of starch results from alkaline or acid hydrolysis, cross-linking, and oxidation, as well as the introduction of functional groups, like acetyl-, carboxyl-, ethyl-, or octenyl-, to starch –OH groups (Compart et al., 2023).

Physical modifications through thermal processes, such as pre-gelatinization, extrusion, heat-moisturizing, and microwaving, or non-thermal processes, including ultra-high pressure, ultrasonication, and pulse electric field exposure, may interrupt or modify granular starch size or arrangement. Starch modifications by α-amylases, β-amylases, isoamylases, and cyclodextrin glycosyltransferase result in the hydrolysis of alpha-1,4 and alpha-1,6 linkages (Compart et al., 2023). It is important to highlight that the use of modified starches in foods has been considered safe according to the Joint FAO/WHO Expert Committee on Food Additives (JECFA, 2016), which designates them as an acceptable daily intake (ADI) as 'not specified'. Regarding this, the EFSA Panel on Food Additives and Nutrient Sources added to Food (ANS) stated a scientific opinion on the reassessment of safety issues (EFSA, 2017) and reports that modified starches are not under genotoxic concern based on *in silico* analysis. For human evaluations, no treatment-related effects were observed using rats fed with very high levels of modified starches (>31,000 mg/kg per day), and doses of 25,000 mg/kg were well tolerated by humans. Using a read-across approach, the data on long-term toxicity and carcinogenicity of modified starches are considered sufficient, and there are no safety concerns within the reported levels (EFSA, 2017).

Eleven modified starches, included as carriers, are approved as food additives in the European Union list in *quantum satis* maximum level (Regulation EC 1333/2008), including oxidized starch (E 1404), monostarch phosphate (E 1410), distarch phosphate (E 1412), phosphated distarch phosphate (E 1413), acetylated distarch phosphate (E 1414), acetylated starch (E 1420), acetylated distarch adipate (E 1422), hydroxypropyl starch (E 1440), hydroxypropyl distarch phosphate (E 1442), starch sodium octenyl succinate (E 1450), and acetylated oxidized starch (E 1451). However, the Panel stated that sodium octenyl succinate in dietary foods for special medical purposes and special formulas for infants requires additional evaluation data on the potential health effects, as well as the oxidized, monostarch phosphate, distarch phosphate, phosphate distarch phosphate, acetylated distarch phosphate, acetylated, sodium octenyl succinate, and acetylated oxidized starches in dietary foods for babies and young children for special medical purposes.

Modified starches such as β-cyclodextrin, maltodextrin, and octenyl succinic anhydride (OSA) are frequently used in starch-based microcapsules due to their physicochemical properties, which enhance the stability of volatile compounds against light, heat, and oxygen. These starches show emulsifying characteristics, forming an oil-water interface that allows the encapsulation of hydrophobic compounds, increasing their solubility in water (Table 1) (Zhao et al., 2023). Starch-based microparticles are found to provide controlled release, stability to gastrointestinal environments and preserve biological activity of encapsulated bioactives (Witczak et al., 2015; Egharevba, 2019; Liang et al., 2023).

Hydrolyzed starch (2.5% hydrolysis grade) has been shown to encapsulate 76% of carotenoids from red araçá efficiently (*Psidium cattleianum* Sabine) extracts when employed at a 1:1 wall-core ratio (w/w) through the spray-drying technique. It also improves total phenolic, carotenoid, and anthocyanin retention and maintains antiradical activities of loaded compounds as demonstrated by 2,2-diphenyl-1-picrylhydrazyl (DPPH•) and 2,2′-azinobis (3-ethylbenzothiazoline-6-sulphonate (ABTS•+) assays when compared to starch combinations with other coating agents, such as taro gum and arabic gum. The half-life of the microencapsulated compounds ranged from 23 to 37 days following microencapsulation at a 1:1 (w/w) starch: taro gum ratio (Rosário et al., 2020).

Modified OSA-starch microparticles were found to retain 67% anthocyanins from jussara (*Euterpe edulis* Martius) pulp, promoting 83% water solubility when encapsulated at a 1:0.5 core-to-wall ratio (w/w) by spray drying, superior to inulin and maltodextrin combination employed as a wall component, demonstrating the potential of modified starch alone to retain and confer polyphenol water solubility (Lacerda et al., 2016). However, a wall material mixture comprising 66% OSA-starch, 16 % maltodextrin, and 16% inulin appears to be the ideal composition to ensure anthocyanin stability and antioxidant polyphenol power following 38 days under unfavorable storage conditions, 50 °C, and light exposure (Lacerda et al., 2016). Encapsulation efficiencies >80% were achieved employing porous corn starch as the coating material for curcumin and resveratrol, both solubilized in ethanol (1:10, w/w). However, saturation of the porous starch surface area can restrict polyphenol loading, affecting encapsulation efficiency and requiring additional experimental data to better employ this microencapsulation system (Wahab & Janaswamy, 2024).

Starch derived from taro has also been modified by OSA esterification to coat phytosterol and phenolic compounds from pomegranate (*Punica granatum L.*) seed oil employing β-cyclodextrin (β-CD) as an emulsifier. The OSA-starch microparticles produced by spray drying at 15% solids concentration and at a 1:3 core-to-wall ratio were shown to encapsulate 61% of total phytosterols and phenolic compounds. The microencapsulated pomegranate oil released 6.63% and 49.8% of bioactive compounds under simulated gastric and intestinal conditions, respectively (Cortez-Trejo et al., 2021). The bioaccessibility of polyphenols coated in native taro starch by spray drying was higher, especially at pH 8, where these compounds can become unstable. However, 12.3% of the encapsulated polyphenols remained stable following intestinal digestion compared to only 2.7% of unencapsulated compounds (Rosales-Chimal et al., 2022).

A microencapsulated and ready-to-eat beetroot (*Beta vulgaris L.)* soup was formulated using native starch to coat the beetroot matrix, which is rich in polyphenols, nitrate, and minerals like potassium, magnesium, zinc, and phosphorus. The encapsulation in starch by freeze drying formed small and spherical particles capable of encapsulating 55% of total betalain content, the main bioactive beetroot compound, at a 2:1 starch-to-core ratio (w/w) (Trindade et al., 2023; Patent BR1020230151965). These findings indicate that starch is effective for microencapsulating polyphenols and pigments and could be applied to developing new foods that confer health benefits. The studies described here demonstrated that no single starch microparticle composition can be considered superior for active compounds delivery, since each formulation offers specific advantages to the bioactive compound microencapsulated, such as higher retention, solubility, and stability, because the microencapsulation depends on the variability of methodological approaches used to entrap the bioactive molecules. Therefore, it is worthwhile to prioritize comparative evaluation under standardized conditions between starches from different sources, including the native and modified ones, exploring combinations with other polysaccharides and their physicochemical behaviors under *in vivo* conditions.

## 2.2 Maltodextrin

Maltodextrin is a polysaccharide obtained through acid or enzymatic starch hydrolysis applied in a controlled way (Figure 1), generating D-glucose polymers linked by α-1,4 and α-1,6 glycosidic bonds, representing about 3% glucose and 7% maltose (w/w). As hydrolysis extends along the starch chain, maltodextrin polymers exhibiting variable molecular weights and reducing-end group contents can be generated (Xiao et al., 2022)

The polymers generated are classified by their dextrose equivalents (DE), indicating how much of the starch structure has been hydrolyzed into simpler sugars (Chronakis, 1998). Maltodextrins DE range from 3 to 20, resulting in polymers displaying distinct physical and chemical properties, such as water solubility, freezing point, and viscosity. High DE maltodextrins, between 16 and 20, composing microcapsules, provide good stability to volatile compounds in foods, due to smoother surfaces, smaller diameters, the presence of vacuoles, and more –OH groups interacting by hydrogen bonds (Wang & Wang, 2000; Xiao et al., 2022).

Maltodextrin at high concentrations forms a network of double helical chains and long-chain aggregates. As a hygroscopic and water-soluble substance, maltodextrin can assume an elastic helical structure with a hydrophobic core, which allows for its complexation with bioactive compounds (Xiao et al., 2022), displaying desirable properties for encapsulation such as thermal and acid stabilities, freezing resistance, ability to prevent core oxidation, and high-water solubility. The use of maltodextrin as a food additive increases solids concentrations and reduces hygroscopicity. Maltodextrin also increases the glass transition temperature (Tg) of food products, resulting in greater thermal and oxidative stability of volatile compounds and microencapsulation efficiency. Even though maltodextrin alone is very effective as a coating material, it can also be combined with Arabic gum, modified starch, or inulin to improve its limited emulsifying capacity, resulting in more stable microencapsulating emulsions (Xiao et al., 2022).

The remarkable maltodextrin encapsulation efficiencies for polyphenols and pigments are shown in Table 1. Annatto (*Bixa orellana L.*) extracts were encapsulated with 86% efficiency, reaching 97% solubility, higher bixin and norbixin contents, low hygroscopicity, and a 60-day microcapsule storage stability (Shridar et al., 2024). A similar microencapsulation efficiency was reported when maltodextrin was employed to prepare freeze-dried beetroot *(Beta vulgaris L.)* microparticles, since 88% of betalains and higher concentrations of phenolic acids, catechin, epicatechin, and myricetin were trapped (Mkhari et al., 2023).

The encapsulation of black carrot (*Daucus carota L.*) juice in DE 20 maltodextrins through spray drying and freeze-drying techniques resulted in a 98% encapsulation efficiency, reaching an anthocyanin content of 1461.23 mg/100 g. The high encapsulation efficiency is probably due to the hydrolyzed starch in maltodextrin, which stabilizes anthocyanins due to the greater amount of –OH available for interaction, as already mentioned. As expected, encapsulated microparticles displayed higher antioxidant activity (329.40 mmol Trolox/g), surpassing the non-encapsulated juice. The produced microcapsules were 96% soluble in water, reinforcing the importance of microencapsulation regarding hydrophobic active food components (Murali et al., 2014). The greater antioxidant capacity of encapsulated anthocyanins can also be justified by the ability of maltodextrin to form a protective film on the surface of the microcapsules, and this can preserve the bioactivity of anthocyanins (Xiao et al., 2022).

On the other hand, maltodextrins with a lower degree of hydrolysis, as found in DE10 maltodextrins, were able to encapsulate 87% of stevia extract (*Stevia rebaudiana*) by spray drying, which is higher than the 76% found for DE19 maltodextrin. The authors hypothesized that maltodextrins with greater DE have many free –OH that do not interact sufficiently with each other in the polymer chain, which can result in greater exposure of the active nucleus and affect the encapsulation (Zorzenon et al., 2020).

Given these limitations, future experiments should focus on the correlation between the degree of maltodextrin (DE) hydrolysis and the structural and functional stability of the microparticles, the standardization of processing parameters to maximize microencapsulation efficiency, the combination with distinct polysaccharides, the kinetics of compound release, and the evaluation of the biological effects intended.

## 2.3 Alginate

Alginate is a hydrophilic and anionic polysaccharide formed by β-D-mannuronic (M) and α-L-guluronic acid (G) residues arranged in a linear chain through 1,4-glycosidic bonds with specific regions termed M-blocks, G-blocks, and MG-blocks (Figure 2), in which M and G contents are variable according to the source (Yerramathi et al., 2023; Puscaselu et al., 2020; Bannikova et al., 2018). The cell walls of *Phaeophyceae* brown algae are the source of alginate extraction, which exists as mixtures of calcium, sodium, potassium, and magnesium salts (Flórez-Fernández et al., 2019; Yerramathi et al., 2023).

Alginate has great application in the food industry due to its ability to form a stable gel-like structure through ionic crosslinking of α-L-glucuronic carboxylate groups in the presence of divalent ions, such as calcium chloride, providing consistency to food preparations and stability to volatile compounds. The gelling property of alginate enables the development of edible films and stable formulations employed in commercial food products and drug delivery systems (Bi et al., 2022). The physicochemical properties of alginate have stimulated its employment for therapeutic purposes as an oral delivery vehicle due to its stability in the gastric environment and gradual dissolution under alkaline fluids from the small intestine, making this polysaccharide a suitable vehicle for bioactive compounds stability in oral administration (Table 1) (Sandberg et al., 1994).

Alginate can be combined with chitosan through ionic cross-linking between alginate carboxylate moieties and protonated amines of chitosan, forming a three-dimensional entrapment matrix that displays valuable physicochemical characteristics for active microencapsulation carriers (Lawrie et al., 2007). For example, alginate, chitosan, and calcium chloride microparticles have been employed to load phenolic acids obtained from orange peels. These polysaccharides generate 252 µm carriers able to encapsulate 89% of active compounds, in turn improving the antioxidant capacity evaluated by $IC_{50}$ values, with 55% of all compounds released after two h in a simulated-gastric fluid, and 82.5% within 24h in intestinal fluid (Savic et al., 2022). Alginate-chitosan microparticles obtained by extrusion have also been employed to encapsulate a black locust flower extract, with a 92% efficiency and a sustainable release reaching about half the amount present in gastric and intestinal simulated fluids after 2 h and 6 h, respectively. In both assays, antioxidant activity maintenance following gastrointestinal digestion was a result of polyphenol, organic acid, and saponin stabilization from the microcapsule-loaded extracts alongside the synergistic antioxidant power inherent to both chitosan and alginate molecules (Boškov et al., 2024).

Alginate as a microencapsulation coating protects bioactive compound structures in physiological fluids, ensuring their delivery, and guarantees volatile compound stability during storage under harsh environmental conditions concerning light, oxygen, humidity, and temperature. In this sense, flavonoids from herbal galactagogue were stable at 4ºC and 25ºC for 120 days after being loaded in a water/oil emulsion when covered by sodium alginate microparticles and encapsulated in chitosan. The optimal conditions for flavonoid stability were determined as 1.49% sodium alginate, 0.84% calcium chloride, 1.58% herbal extract, and loading alginate-microemulsions in 1% chitosan. These carrier systems generated 123 µm-sized microparticles capable of encapsulating 77% of all flavonoids and promoting controlled simulated gastrointestinal release for 24h (Khorshidian et al., 2019). Interestingly, alginate-based microbeads can also be employed as β-carotene nanoemulsion stabilizers for 12 days at 55 ºC. Alginate microparticles protects lipid-derived structures from lipase activity, conferring up to 87% stability in a simulated gastrointestinal environment by encapsulating nanoemulsions in 0.5% alginate crosslinked with 10% calcium chloride at a 1:1 ratio (core-to-matrix) (Zhang et al., 2016). Other polyphenols, such as catechin, astaxanthin, and quercetin, are efficiently encapsulated in alginate, alone or combined with chitosan, resulting in an improved release and chemical stability under physicochemical conditions (Kim et al., 2016; Niizawa et al., 2025; Frent et al., 2022).

In general, although under simulated conditions, the current scientific evidence shows that, through electrostatic interactions, alginate, mainly combined with oppositely charged polysaccharides, displays desirable behavior for carriers intended for the oral delivery of nutraceuticals, as it has exhibited high gastric resistance and sustained intestinal dissolution, with the potential to increase the bioaccessibility of orally administered active compounds.

## 2.4 Pectin

Pectin is a complex heteropolysaccharide formed by at least 17 kinds of monosaccharides, including D-galacturonic acid as the most prevalent, comprising nearly 70% of monomers, D-galactose, L-arabinose, among others (Figure 3) (Roy et al., 2023; Roman-Benn et al., 2023; Barrera-Chamorro et al., 2025). Pectin refers to a family of polysaccharides from primary cell walls and middle lamella, which contribute to plant maintenance and present structural variations composed mainly of homogalacturonan, known as the 'smooth' region, and rhamnogalacturonan II, the 'hairy' region. The backbone is composed of D-galacturonic acid units linked by α-1,4 glycosidic bonds (Dranca & Oroian, 2018)

Pectin is extracted from agro-industrial residues, *i.e.*, sugar beet pulp, citrus peel, and apple pomace, and employed as a food additive thickener, stabilizer, and gelatinizer, due to its rheological and textural properties that improve sensorial food product features (Berlowska et al., 2018; Chan et al., 2017). The physicochemical properties of pectin are influenced by its structural variability, mainly concerning the degree of esterification (DE), defined as the percentage of carboxyl (–COOH) groups esterified with methyl groups, in which DE >50% is considered high methoxylated, and DE <50% is low methoxylated.

Pectin is widely employed in the pharmaceutical industry as a polymeric encapsulant matrix for drug entrapment, contributing to drug stability and delivery, since its high methoxylation degree enables pectin to interact with hydrophobic molecules, improving drug loading and providing controlled release of the active compound (Rehman et al., 2019; Galvez-Jiron et al., 2025; Cacicedo et al., 2018). The formation of pectin carriers is underlined by ionic interactions between its –COOH groups in the presence of divalent ions, such as $Zn^{2+}$. The divalent crosslinking controls the release of encapsulated compounds and, alongside polymers like alginate, promotes the formation of an "egg box" structure, making microstructures very resistant due to the linkage between galacturonate and guluronate blocks (Oh et al., 2020).

Several reports demonstrated pectin's structural properties and physicochemical interactions with loaded compounds, which could result in stable complexes able to protect and transport bioactive compounds at environmental and physiological conditions, being therefore reliable for application to food matrices and nutraceutical formulations.

Pectin microparticles designed for resveratrol entrapment were synthesized by extrusion and crosslinking with $Zn^{2+}$, forming 900–950 µm particles capable of retaining 94% of polyphenols and promoting 90 days of storage stability at 25ºC and 40ºC. The pectin microparticles conferred protection to resveratrol greater than 90% in simulated gastric conditions (pH 1.2); in other words, only a small amount of resveratrol is released after 2 h in gastric digestion fluid (Das et al., 2011). Similarly, increased gastrointestinal resistance of resveratrol can be achieved through its entrapment in pectin/alginate blend microparticles (polymer ratio 2:1 w/w) formed by ionotropic gelation with calcium chloride, followed by atomization in a peristaltic pump. This combination stabilizes resveratrol for 2h upon gastric fluid (pH 1.2), releasing less than 10% of the polyphenol (Gartziandia et al., 2018).

Pectin combined with alginate (55:45 w/w polymer ratio), calcium chloride can generate 180 µm microparticles that provide sustained gallic acid release and stability, resulting in 40% and 30% releases at pH 1.8 and pH 6.5, mimicking gastric and intestinal fluids, respectively, and complete release after 4 h, a timeframe close to human digestion (Vallejo-Castillo et al., 2020). Gallic acid microencapsulation of up to 89% efficiency was obtained by air drying, forming 1300 µm pectin-alginate microparticles (1:1 w/w polymer ratio), with cumulative release percentage ranges of 15–35%, 55–85%, and 50–85% at pH 4, pH 7, and pH 10, respectively (Nájera-Martínez et al., 2023).

In short, the physicochemical behavior and compatibility with food components allow the development of pectin carriers based on intermolecular linkages, such as crosslinking and hydrogen bonds with calcium and alginate. These interactions generate a stable network microstructure around dispersed bioactive compounds in the nucleus with no chemical modifications, promoting high gastrointestinal resistance, controlled release, and bioactive content stability, enabling them to be absorbed in the gastrointestinal mucosa. Crosslinking and polymer interactions achieved by using pectins enhance the protection and can improve the bioavailability of bioactive compounds.

### 2.5 Inulin

Inulin is a naturally occurring polymer belonging to the fructan subgroup, comprising β-d-fructose units joined by β-(2-1) glycosidic bonds, typically displaying a terminal glucose residue. Inulin can exhibit linear or slightly branched structures with a degree of polymerization (DP) ranging from 2 to 60 monomers, categorized as short-chain inulin (DP < 10) and long-chain inulin (DP > 10), each of them exhibiting variable physicochemical characteristics (Figure 4). The DP of inulin varies according to plant source, cultivar, environmental conditions, harvest time, and post-harvest processing. These DP variabilities underlie differences in physicochemical behavior, affecting solubility, gelation, and overall functionality (Mensink et al., 2015; Ni et al., 2019; Mudannayake et al., 2022; Gruskiene et al., 2023).

Inulin is widely distributed among various plant species, with the richest sources comprising members from the Asteraceae family, particularly chicory (*Cichorium intybus L.*), Jerusalem artichoke (*Helianthus tuberosus*), dahlia (*Dahlia spp.*), and agave (*Agave spp.*) roots. Chicory and Jerusalem artichoke are the primary inulin sources used in the food industry, as they are easy to process (Mudannayake et al., 2022; Gruskiene et al., 2023).

Inulin has a well-documented prebiotic effect and exhibits a broad range of functionalities, *i.e.*, fat-replacer, immunological system modulation, calcium absorption, and several other metabolic processes that contribute to gut health and risk of metabolic disorders (Letexier et al., 2003; Seifert & Watzl, 2007; Nair et al., 2010; Lavanda et al., 2011; Gupta et al., 2019). Due to its chemical structure versatility, inulin can also be employed as a drug carrier in food and nutraceutical formulations. Inulin is resistant to the gastric fluid and can form protective wall matrices, enhancing the stability of loaded bioactive compounds against oxidation, pH, and temperature variations (Gupta et al., 2019; Setyaningsih et al., 2022; Akram et al., 2024; Gruskiene et al., 2023; Zhang et al., 2024).

Short- and long-chain inulin serve as effective microencapsulating agents, although encapsulation efficiency depends on the physicochemical properties of the entrapped compounds and their interaction with the inulin matrix. Short-chain inulin is particularly suitable for encapsulating hydrophilic polyphenols such as flavonoids, catechins, epicatechins, and rutin or natural pigments, including anthocyanins and betalains, due to its lower molecular weight and higher –OH content, making it highly soluble (Ayvaz et al., 2022) (Table 1). Its high number of hydrophilic groups allows for hydrogen bonding, improving encapsulated compound dispersibility, bioaccessibility, and solubility in aqueous fluids, and it is ideal for water-based mixtures and functional foods. In contrast, long-chain inulin exhibits lower water solubility but superior gelling and emulsifying properties, forming a structured polymeric network comprising a protective barrier for loaded compounds. This denser matrix protects both lipophilic and amphiphilic compounds, such as resveratrol, curcumin, and carotenoids, from oxidation, light, and temperature degradation, making it useful for their storage and also for controlled and sustained release of the compounds (Mensink et al., 2015; Liu et al., 2022; Gruskiene et al., 2023). Although these contrasting properties expand the range of applications, they underscore the need to carefully select the inulin type based on the polarity of the target bioactive. Otherwise, poor solubility or weak structural retention can reduce encapsulation performance in real food systems (Mensink et al., 2015; Liu et al., 2022).

The encapsulation efficiency of inulin can also be enhanced through combinations with maltodextrin, alginate, and proteins, forming hybrid structures that improve the retention and stability of loaded compounds (Liu et al., 2022; Gruskiene et al., 2023). Although hybrid systems enhance stability, they also complicate the standardization of wall formulations, since performance will depend on inulin chain length and the physicochemical interactions with co-polymers. Such variability challenges the reproducibility across batches and processing conditions (Liu et al., 2022; Gruskiene et al., 2023).

The encapsulation of polyphenols and pigments into inulin microparticles, combined with other encapsulating agents, has demonstrated benefits concerning the stability, thermal resistance, and bioavailability of these compounds. Olive leaf (*Olea europaea*) extracts encapsulated in inulin by spray-drying, for example, were shown to improve the intestinal availability and stability of hydroxytyrosol and oleuropein, two phenolic compounds displaying strong antioxidant and anti-inflammatory properties, but highly susceptible to enzymatic hydrolysis and acidic stomach degradation. Encapsulation achieved ~80% retention rates and protected these compounds while enabling their controlled release in the intestinal lumen. This protective effect was attributed to hydrogen bonding and hydrophobic interactions between inulin and polyphenols, later confirmed by FTIR analyses, which indicated structural stabilization and protection against premature polyphenol degradation (Duque-Soto et al., 2023). Other studies have confirmed that inulin coatings also protect oleuropein in spray-dried microparticles displaying spherical morphology (~10 µm), an encapsulation efficiency of 78 %, and a recovery rate of 70.5 %. Up to 60% of oleuropein was released under simulated gastrointestinal digestion, suggesting that inulin facilitates colonic delivery through fermentation. Additional analyses indicated superior oleuropein retention in baked matrices (~77 mg/100 mg) compared to boiled ones (~63 mg/100 mg), highlighting the protective effect (Pacheco et al., 2017). Although inulin has proven its effectiveness in protecting phenolic compounds such as oleuropein and hydroxytyrosol, comparative studies revealed that maltodextrin or modified starch may achieve higher retention for some pigments, such as betalains. This indicates that the choice of encapsulating agent must be aligned with the specific chemical nature of the target bioactive compound.

Inulin can also be employed as an alternative to modified starch in the encapsulation of oregano extracts (*Origanum vulgare L.*). A high encapsulation efficiency of up to 66 % and greater thymol retention, reaching 84 %, and enhanced thermal stability up to 220°C, were noted. Scanning electron microscopy analyses confirmed a uniform and spherical microcapsule morphology, highlighting the suitability of inulin as a prebiotic encapsulant concerning lipophilic compounds (Zabot et al., 2016).

Furthermore, quercetin encapsulation employing a combination of inulin, alginate, and chitosan to target colonic release has also been successful. The presence of inulin improved structural microspheres integrity by filling polymeric pores, which led to enhanced gel strength and controlled quercetin release. *In vitro* simulated digestion experiments demonstrated that the encapsulated quercetin obtained by freeze-drying displayed a 50% bioavailability increase compared to free quercetin, with encapsulation efficiencies reaching between 70-85%, followed by sustained release up to 24 h. Furthermore, FTIR analyses confirmed strong hydrogen bonding interactions between quercetin and wall materials (Liu et al, 2022).

Another study demonstrated that polyphenol microencapsulation by spray-drying of blackcurrant (*Ribes nigrum L.*) compounds employing different ratios of inulin and maltodextrins displayed better phenolic compound retention and higher antioxidant activity maintenance after 12 months of storage at 8°C and 25°C. Inulin was noted as effective in protecting the microencapsulated compounds against oxidative degradation and in prolonging their antioxidant activity (Bakowska-Barczak & Kolodziejczyk, 2010). Similarly, gallic acid encapsulation into inulin resulted in spherical microparticles with smooth surfaces after spray-drying, with an encapsulation efficiency of 83% in native inulin, higher than when employing native starch, indicating better polymer-gallic acid interaction. The bioaccessibility of encapsulated gallic acid was determined in hydrophilic media, with a quick release timeframe (< 9 h), indicating a promising use of microencapsulated gallic acid-inulin as a functional ingredient in dry mixtures or in instant foods (Robert et al., 2012).

Several other studies have also demonstrated that inulin microencapsulation contributes to higher pigment encapsulation efficiency. This improves thermal pigment stability and enhances antioxidant activity retention while optimizing and controlling their release when exposed to simulated digestive tract conditions. In this sense, the microencapsulation of natural pigments, such as betalains, anthocyanins, and carotenoids within inulin or a combination of biopolymers has proven effective regarding thermal stability and color retention. Betalains loaded in inulin microparticles obtained by spray drying, for example, resulted in a 58.4% encapsulation efficiency. When added to sorbets, betalains-loaded microcapsules significantly improved sorbet color stability over six months at -18°C, reinforcing the role of this polysaccharide in coating and protecting pigments in frozen food matrices (Omae et al., 2017). Similarly, betalains encapsulation using different wall materials, such as inulin alone, inulin combined with maltodextrin (IN-MD), and inulin combined with a whey protein isolate (IN-WPI) by spray-drying resulted in over 88% betalains retention. While the different polysaccharide coatings did not significantly affect retention, the IN-WPI mixture exhibited the highest powder stability, suggesting that inulin combined with proteins protects betalains (Carmo et al., 2017). However, when betalains and other related pigments (betacyanins and betaxanthins) were encapsulated by freeze drying in inulin or maltodextrin as carrier agents, maltodextrin-based microparticles retained the highest betalains content (15.72 mg/100 g) compared to inulin (10.11 mg/100 g). This lower retention has been attributed to the relatively low glass transition temperature ($T_g \approx 27$ °C) of inulin-based powders, which makes them more prone to collapse and degrade under warm or humid storage conditions. In contrast, maltodextrin systems ($T_g \approx 60$ °C) display greater stability. Therefore, inulin may perform better in frozen or refrigerated products, but its application in thermally processed foods often requires blending with higher-Tg polymers to ensure structural stability (Flores-Mancha et al., 2020). An inulin-alginate combination has been tested by extrusion, demonstrating that increasing inulin from 10% to 20% (w/w) improves encapsulation efficiency from 79% to 89%, with total betacyanin contents reaching 4.21 mg/g, and then reinforcing the potential of inulin-alginate systems for pigment stabilization (Azevedo & Noreña, 2021). Inulin can be considered for betalain encapsulation, although its performance is strongly dependent on the chosen wall combination, and optimization with proteins or maltodextrin is often required to achieve consistent stability across storage conditions.

The encapsulation of anthocyanin into inulin microparticles or mixtures of inulin and other polysaccharides has been shown to protect and enhance the bioavailability of these natural pigments. For example, the encapsulation of anthocyanins obtained from maqui (*Aristotelia chilensis*) juice tested in both inulin and alginate by spray-drying resulted in high encapsulation efficiencies, reaching 78.6% when inulin was employed as the coating agent. The bioaccessibility enhancement of encapsulated anthocyanins confirmed their preservation and delivery under gastrointestinal physicochemical conditions (Fredes et al., 2018).

Anthocyanins from a chokeberry (*Aronia melanocarpa*) extract were microencapsulated by spray drying into an inulin and maltodextrin mixture, resulting in up to 90 % water solubility, an 88 % encapsulation efficiency, and moderate anthocyanin degradation (10.38%) under 7 days of storage, exposed to light and room temperature (Pieczykolan & Kurek, 2019).

Encapsulation methods affect anthocyanin retention and have been assessed by comparing spray-drying and freeze-drying in the encapsulation of a roselle (*Hibiscus sabdariffa*) extract employing an inulin and maltodextrin mixture (Nguyen et al., 2021). The best performance was obtained by spray-drying when considering total polyphenol content and anthocyanin retention, as well as higher antioxidant activity preservation. Anthocyanins from a mangosteen (*Garcinia mangostana*) pericarp extract were encapsulated in inulin and maltodextrin by spray-drying. The microencapsulated compounds reached 80.94% encapsulation efficiency and adequate anthocyanin preservation, with an anthocyanin half-life (t1/2) of 3.16 days when incorporated into yogurt and stored at 4°C (Sakulnarmrat et al., 2022). These results reinforce the potential of inulin for the microencapsulation of anthocyanins, but also highlight its dependency on process parameters, since differences between spray- and freeze-drying directly affect stability and retention of pigments, making the method selection a critical factor for successful application.

In addition, carotenoids have also been shown to benefit from inulin encapsulation. The encapsulation of β-carotene extracted from mango peel (*Mangifera indica*) by spray drying coated by inulin and fructooligosaccharides (FOS), for instance, reached 47% bioaccessibility compared to formulations without these prebiotics, which reached just 28-30% FOS. Ultramicroscopy images revealed post-digestion micelle-like formations, facilitating β-carotene solubilization and increasing its uptake in Caco-2 cells, reinforcing inulin role in improving carotenoid absorption (Cabezas-Teran et al., 2022). Similarly, the encapsulation of lycopene obtained from tomato (*Lycopersicon esculentum*) in inulin microparticles synthesized by spray-drying led to a 26% final release of microencapsulated lycopene during simulated intestinal digestion, whereas only 15% of free lycopene was recovered after simulated digestion, suggesting that encapsulation significantly enhances intestinal lycopene delivery and consequent bioaccessibility (Corrêa-Filho et al., 2022). However, as well as the polyphenols and anthocyanins, carotenoid stability in inulin microparticles is highly sensitive to storage temperature and humidity, reinforcing that the formulation must be tailored to the physicochemical fragility of the target compound.

Inulin emerges as a versatile biopolymer with recognized prebiotic activity and effective capacity as a carrier for bioactive compounds. Its structural variations and combinations with other polymers influence positively on its functionality, enabling applications as functional compound and carrier in delivery systems.

## 2.6 Chitosan

Chitosan is an amino polysaccharide derived from the alkaline deacetylation of chitin, a biopolymer found in exoskeletons of crustaceans, fungi cell walls, insect cuticles, and certain microalgae. It consists of a non-branched chain composed of β-(1,4)-linked D-glucosamine (deacetylated units) and N-acetyl-D-glucosamine (acetylated units) (Figure 5). It is a biocompatible, biodegradable, and non-toxic polymer, exhibiting over 60% degree of deacetylation (DD) and molecular weight (Mw) ranging from 50 to 2000 kDa (Iber et al., 2021; Weißpflog et al., 2021; Meyer-Déru et al., 2022; Carrasco-Sandoval et al., 2023).

Chitosan undergoes chemical, enzymatic, and/or physical degradation, generating oligosaccharides with specific biological properties. Moreover, its reactive $–NH_2$ groups enable the synthesis of derivatives displaying enhanced bioactivity, improved water solubility, and antioxidant and anti-inflammatory activities. These physicochemical characteristics enhance the stability and bioavailability of entrapped bioactive molecules while also enabling chitosan formulations as solutions, films, hydrogels, fibers, nanoparticles, and microparticles. The versatility of chitosan associated to its low cost enhances its employment for the development of pharmaceutical, food, agricultural, and environmental products (Gomes et al., 2018; Haghighi et al., 2018; Aranaz et al., 2021; Mawazi et al., 2024; Barbosa-Nuñez et al., 2025).

Physicochemical properties and biological performance of chitosan depends on the degree of deacetylation (DD), defined by the molar fraction of repeating units with containing free amino groups ($–NH_2$), while the degree of acetylation (DA) corresponds to the proportion amount of N-acetylated units, expressed as percentage of total units and the molecular weight (Mw), in addition to the physicochemical conditions such as the ionic strength, pH, temperature, and solvent polarity. All of them influence the ability of chitosan to form polyelectrolyte complexes, resulting in its suitability as an agent for drug delivery, since they determine chitosan solubility, intermolecular interactions, affecting its biological activity. Lower DA results in a high content of protonatable $–NH_2$ groups, which become positively charged ($–NH_3^+$) under acidic conditions (pH < 6.5), enhancing the chitosan cationic character and its solubility. Conversely, high DAs promote stronger hydrogen bonding and hydrophobic interactions due to the presence of N-acetyl-D-glucosamine units, leading to increased crystallinity and reduced solubility (Aranaz et al., 2021; Kou et al, 2020).

Similarly, the Mw significantly affects the impact on chitosan functionality. Low Mw chitosan displays enhanced gastrointestinal tract permeability and absorption in the gastrointestinal tract by crossing its epithelial barriers. Conversely, high Mw chitosan contains a higher density of cationic charges to chitosan molecules, which tend to exhibit stronger interactions with cell membranes, promoting aggregation and potentially increasing cytotoxicity. These physicochemical characteristics must be controlled for the use of chitosan as carriers for microencapsulation, facilitating chitosan water solubility at acidic pH, the formation of polyelectrolyte complexes with negatively charged bioactive compounds, enabling their encapsulation and controlled release. (Meyer-Déru et al., 2022; Yadav et al., 2022; Carrasco-Sandoval et al., 2023; Harugade et al, 2023).

The cationic nature of chitosan allows its interaction with various biomolecules via electrostatic attraction, hydrogen bonding, and hydrophobic interactions, depending on the chemical nature of the compounds to be encapsulated (Wang et al., 2006; Aranaz et al., 2021).

Besides acting as an encapsulating agent, chitosan provides mucoadhesion, enhances permeability, and stabilizes bioactive compounds and their controlled release (Raza et al., 2020; Aranaz et al., 2021; Yadav et al., 2022) (Table 1). The mucoadhesive properties of chitosan are particularly valuable in oncology, as they enhance the therapeutic effectiveness of anticancer compounds by improving drug retention in tumor sites. Moreover, chitosan exhibits natural antibacterial properties contributing to reduce post-treatment infections, further supporting its biomedical relevance (Harugade et al., 2023; Wei et al., 2014; Fan et al., 2019; Freitas et al., 2024).

Another chitosan advantage consists on its ability to co-encapsulate bioactive compounds alongside other encapsulating biopolymers, forming more stable and efficient delivery systems. Chitosan combined with proteins, such as the whey protein isolates, or other polysaccharides, including inulin, alginate, and pectin, or structured lipids, can enhance the stability, protection, and bioavailability of the loaded bioactive compounds. Chitosan-alginate nanoparticles were noted as improving the protection and controlled release of folic acid and vitamin E (Tan et al., 2024). Similarly, the co-encapsulation of phenolic compounds employing chitosan and whey protein as wall materials resulted in greater capsule resistance to gastric digestion and, at the same time, enhanced intestinal absorption in *in vitro* models (Kasapoğlu et al., 2024).

The co-microencapsulation of blackcurrant (*Ribes nigrum*) anthocyanins through freeze-drying in a multi-matrix composed of chitosan, whey protein isolate (WPI), and inulin produced a complex microparticle structure with aggregates and spherosomes ranging from 5 to 20 µm, where anthocyanins were coated by the biopolymeric matrix at a 95% efficiency, conferring resistance to gastric fluids and allowing for 94% release after 2h at the intestine. Anthocyanins cargo load can be maintained for 90 days at 4°C due to electrostatic interactions between the positively charged chitosan and negatively charged anthocyanins, as indicated by ZP measurements, hydrogen bonding between anthocyanin –OH and $–NH_2$ and –C=O groups between chitosan/WPI, and hydrophobic interactions with WPI and inulin, reducing the exposure to oxidation and/or moisture. These chemical interactions, along with physical entrapment, ensured gastric resistance and controlled intestinal release. The powdered microcapsules was shown to inhibit enzymes involved in carbohydrate metabolism, such as α-amylase and α-glucosidase, by 87% and 37%, respectively, suggesting a potential antidiabetic effect. Even when incorporated into yogurt, anthocyanins remained stable in food matrices for 21 days at 4°C, ensuring their gradual release (Enache et al., 2020).

An aqueous garlic extract encapsulated into chitosan and WPI, at a mass ratio of 0.2:1 (w/w) by spray drying methodology ensured strong electrostatic interactions, charge neutralization, and stable coacervate formation. The spherical microparticles retained phenolic compounds near 60%, increasing thermal stability and solubility in water (76%–94%). However, the high hygroscopicity of coating material, from 21% to 28%, suggests the need for moisture-resistant packaging (Gomes et al., 2015; Tavares & Noreña, 2018).

The microencapsulation of a polyphenol extract from apple pomace into a chitosan-fish gelatin produced by freeze-drying retained over 80% of phenolic compounds in lamellar and vesicle-like structures, with a progressive reduction in the pore size as the extract concentration increased. Strong electrostatic interactions occur between the negatively charged polyphenol compounds and the cationic chitosan-fish gelatin (Moradi et al., 2024).

Glutaraldehyde-crosslinked chitosan microparticles is suitable for the encapsulation of polyphenols from *Thymus serpyllum* by emulsion crosslinking, in which chitosan concentrations ranged from 1.5% to 3% (w/v) and the glutaraldehyde-to-chitosan mass ratio varied between 0.15 and 1.20, tailoring the physicochemical properties of microparticles. Interactions between the polyphenol compounds and the chitosan matrix took place mainly through hydrogen bonding between the phenolic compound –OH groups and the $-NH_2$ and –OH chitosan groups, as demonstrated by X-ray diffraction analysis, which showed that chitosan crystallinity increased following glutaraldehyde crosslinking and polyphenol compound encapsulation (Trifkovic et al., 2015).

Blueberry anthocyanins were also successfully encapsulated in chitosan microparticles crosslinked with either cellulose nanocrystals or sodium tripolyphosphate (TPP) at pH 7.4. The coating employing chitosan-cellulose resulted in particles of about 65 nm, with a higher yield of 6.9 g. Anthocyanin recovery reached 94% retention efficiency, with a favorable distribution within the matrix (Wang et al., 2016).

Chitosan represents a multifunctional biopolymer with remarkable physicochemical and biological properties that support its wide applicability in food, pharmaceutical, and biomedical fields. Its cationic nature, tunable molecular weight, and degree of deacetylation enable versatile interactions with diverse bioactive compounds, facilitating encapsulation, protection, and controlled release.

## 2.7 Gum Arabic

Gum Arabic is a complex and highly branched polysaccharide with a molecular weight ranging from $2.5 \times 10^5$ to $1.0 \times 10^6$, containing magnesium, calcium, and potassium salts. The main chain is formed by β-D-galactopyranosyl units linked by 1,3-linkages, and side chains containing different arabinofuranose, galactopyranose, rhamnopyranose, and uronic acid (glucuronic or galacturonic) units linked by 1,6-glycosidic bonds. Uronic acid and galactose are found in β-D form, whereas arabinose and rhamnose are observed in α-L form (Figure 6). Gum

Arabic structure contains small amounts of amino acids (≈2.25% dry/weight), mainly hydroxyproline, aspartic acid, proline, and serine (Nie et al., 2012; Sanchez et al., 2017; Prasad et al., 2022).

Gum Arabic is the oldest and most well-known exudate gum produced by Acacia senegal, a multipurpose tree with an important environmental and sociological role in Savanna countries in the African continent. This tree can also be found in Pakistan, Oman, and India in well-drained, deep, and sandy soils in warm sub-desert type climates (Sanchez et al., 2017; Prasad et al., 2022).

Freshly collected gum Arabic displays low moisture content (≈10-15%) and a considerable number of impurities like bark, leaves, soil, and sand, requiring processing before marketing. The first and most important gum processing comprises sun drying for 5-15 days to lower moisture contents to below 10%, to prevent fungal contamination and bleach gum, improving its overall quality. Subsequently, either cleaning and grading are usually carried out, manually or using specific machines, removing adhered foreign matter by winnowing or hand picking, and finer objects are removed by sieving. Then, gum Arabic is graded based on size, color, and impurities. It can also undergo a mechanical grinding and breaking process following primary processing, called kibbling, converting the gum into nodules of different sizes. Gum Arabic is found commercially in granule, flake, and powder forms, from orange-brown to pale white, acquiring a paler and glassier appearance in the broken or kibbled forms (Sanchez et al., 2017; Prasad et al., 2022).

After processing, gum Arabic becomes an odorless, hard, tasteless, and translucent gum that does not interact with chemical compounds and is easily soluble in water, forming low-viscosity solutions even at high concentrations and acting as a stabilizer for oil-in-water emulsions. It can be used as a thickener, emulsifier, stabilizer, carrier, firming, bulking, or antioxidant employed in food, pharmaceutical, and cosmetic formulations. As a soluble dietary fiber, it is absorbed in the small intestine and passes without being broken down, where substantial daily doses can be ingested with no adverse effects. Because of this, it has been widely employed as an encapsulating agent, generating a stable matrix capable of protecting and prolonging active cores during gastrointestinal digestion (Taheri & Jafari, 2019).

Table 1 lists several studies in which gum Arabic has been used as a carrier to encapsulate phenolic compounds and pigments. Leaf extracts from R. tuberosa L. and T. diversifolia, both rich in phenolic compounds and terpenoids, have been coated by gum Arabic polymers prepared at optimal conditions using 4% (w/v) gum Arabic at pH 5, and stirring for 60 min, followed by freeze-drying. Due to the amphoteric nature of the polysaccharide chain, i.e., molecules containing –COOH groups and –OH groups that play a role in acid-base reactions (pKa of 3.6), gum Arabic stability is directly influenced by pH. At low pH values, the –COOH group can lose a proton ($H^+$), becoming $–COO^–$, forming a negative charge, and at high pH values, the –OH group can gain a proton, forming a positive charge. Therefore, when these groups ionize (gain or lose charge), gum Arabic becomes more soluble in water. Its encapsulation efficiency increases at pH 5, as some gum Arabic acid groups are not fully ionized, resulting in low viscosity and optimal interactions. Gum Arabic microcapsules have antioxidant activity and potential to be used as an adjuvant for the treatment of diabetes mellitus type 2, since they can inhibit alpha-amylase enzyme in vitro (Almayda et al., 2024).

Flavonoids from ponkan peel extracts were microencapsulated by spray-drying using 5% (w/v) soybean oil in an aqueous gum Arabic dispersion (10%, w/v) and/or 2.5% (w/v) soybean oil, 0.5-5% (w/v) whey protein concentrate in an aqueous gum Arabic dispersion (5%, w/v). When 100 mg of ponkan peel extracts were added to gum Arabic-stabilized oil- and whey protein-in water emulsions, microparticle sizes increased 1.0 µm and 1.4 µm, respectively, with a negative zeta-potential observed for all emulsions. However, when the pH increased from 2 to 7 (using 1M HCl or NaOH solution), the average size of the microparticles reached approximately 1 µm, while the droplet diameter measured around 0.87 µm. The pH shifts also led to a marked increase in negative electrical charge from -2.37 mV to -29.07 mV, demonstrating that the zeta potential of the microparticles is influenced by pH. At pH 7, gum Arabic with a stronger negative charge was adsorbed onto the oil-water droplet surfaces, enhancing the stability of the interfacial layer through electrostatic repulsion and effectively preventing droplet flocculation. Furthermore, over a three-month storage period, the total flavonoid content declined in all gum Arabic-stabilized oil-in-water emulsions (with and without whey protein concentrate). Nevertheless, microencapsulation helped maintain the antioxidant activity of the flavonoids across all emulsions, with this activity closely tied to flavonoid concentration. Notably, emulsions stabilized solely with gum Arabic-stabilized oil-in-water emulsions exhibited higher antioxidant activity compared to gum Arabic-stabilized in oil- and whey protein-in water emulsions, likely due to the reduced flavonoid concentration — approximately half— in the whey-based formulations (Hu et al., 2017).

Microencapsulated curcumin employing coconut milk whey and/or gum Arabic by spray-drying at different concentrations (5, 10, and 15%) has retained curcumin with a gradual encapsulation efficiency increasing according to gum Arabic concentrations, and shows mean particle diameters ranging from 22 to 25 μm, providing a rough surface and increased encapsulation area. The gum Arabic added to the curcumin-coconut milk whey solution occupies empty spaces in the carrier matrix, increasing wall material quality and stabilizing this component by reducing oxygen permeability. Smaller particle sizes increased the available surface area for curcumin encapsulation, leading to higher curcumin retention with increasing gum Arabic concentrations. In addition, higher powder solubility was achieved at ~95% for curcumin-coconut milk whey powder (0% gum Arabic), and decreases gradually with gum Arabic concentrations in curcumin-coconut milk whey powder. A combination of encapsulating agents, such as coconut milk, whey, and gum Arabic, may improve powder retention efficiency and characteristics compared to a single carrier material, enhancing the attributes and offering a good alternative for the encapsulation of polyphenols like curcumin. Furthermore, the curcumin degradation rates that are directly proportional to temperature can be decreased by increasing gum Arabic concentrations, resulting in coconut whey protein-gum Arabic complex formation, improving curcumin encapsulation and stabilization (Adsare & Annapure, 2021).

Anthocyanins from barberry (Berberis vulgaris) extract were easily degraded during storage and processing by exposure to heat, light, and oxygen, so their microencapsulation by spray-drying can be optimized employing different polysaccharide combinations, such as maltodextrin and gum Arabic, maltodextrin and gelatin, compared to maltodextrin. The anthocyanin extract microencapsulation into a maltodextrin and gum Arabic at a core/wall ratio of 25% achieved the highest encapsulation efficiency, moisture content, hygroscopicity, and water solubility, reinforcing that a single encapsulating wall material rarely fills all the requirements for efficient microencapsulation. The branched nature and covalent linkages to amino acids of gum Arabic, which are highly linked to the carbohydrate chain, may act as a film-forming agent for the entrapment of bioactive molecules. The flavylium cation of anthocyanins is less vulnerable to nucleophilic attack by water molecules, increasing the stability of the microencapsulated pigments. Furthermore, the complex formed when the flavylium cation of the anthocyanins interacts with dextrins prevents its transformation to other less stable forms (Mahdavi et al., 2016).

Buriti (Mauritia flexuosa) oil pulp, a native Amazon fruit, composed mainly of tocopherols, oleic and palmitic acid, and β-carotene, displays low stability to heat, light, and oxygen. The best encapsulation efficiency and β-carotene retention were observed for the following microencapsulation by freeze-drying of the Buriti oil with 50% gum Arabic/ 50% inulin mix when compared to 25% gum Arabic/ 75% inulin or 75% gum Arabic/ 25% inulin microcapsules. In addition, 50% gum Arabic/ 50% inulin microparticles displayed the highest water solubility when compared to other treatments. Microparticles composed of 75% gum Arabic/ 25% inulin and 50% gum Arabic/ 50% inulin were the smallest, and all formulations presented a negative zeta potential. This can be explained by the increased repulsion between particles, indicating systems that do not tend to agglomerate as much as the surface loads repel and favor stabilization, extending the shelf-life of formulations. Furthermore, an overlap between the stretching of inulin –OH groups (3300 cm-1) and gum Arabic $–NH_2$ and –COOH groups (at 3330 $cm^{-1}$ and 1049 $cm-^{1}$, respectively) in all microparticles obtained was observed through a broadband between 3650 and 3100 $cm^{-1}$. Stretching saturated alkanes near 2925 $cm^{-1}$ and –C=O near 1740 $cm^{-1}$ were also observed, confirming the presence of Buriti oil fatty acids and indicating its incorporation throughout the microparticle structures (De Oliveira et al., 2022).

In short, Gum Arabic demonstrates significant potential as a multifunctional encapsulating agent for phenolic compounds, pigments, and oils, owing to its amphoteric nature, solubility modulation by pH, and film-forming capacity. Its performance and versatility as a drug carrier is further enhanced when combined with other biopolymers, conferring a broad applicability in the development of functional foods and nutraceutical formulations.

## 3. A BRIEF OVERVIEW OF THE PUTATIVE INTERACTIONS BETWEEN POLYSACCHARIDES AND BIOACTIVE COMPOUNDS IN DRUG CARRIER SYSTEMS

Polysaccharides as bioactive carriers can promote health benefits by improving the stability of bioactive compounds and displaying synergic effects through their functional roles on obesity, type 2 diabetes, hypercholesterolemia, and gut microbiota health (Farid et al., 2024;

Lukova et al., 2023; Dharanie et al., 2024). Based on the health benefits, there is a considerable interest in developing polysaccharide microparticles and understanding the chemical interactions underlying bioactive interaction and retention, providing a basis for the choice of a suitable wall material for bioactive delivery (Zhang et al., 2020).

Polysaccharides form delivery systems by binding with bioactive compounds through non-covalent interactions such as hydrogen and ionic bonds, Van der Waals forces, and electrostatic and hydrophobic interactions (Bordenave et al., 2013). On the other hand, covalent linkages between polysaccharides and bioactive compounds are chemical interactions, and usually require conjugation with proteins (Liu et al., 2016). The supramolecular interactions between polysaccharides and bioactive compounds are still underexplored and restricted to polyphenol-polymer chemical interactions, although physical linkage between the polysaccharides and polyphenols or bioactive compounds has been found to enhance the retention of the microencapsulate compound (Xiao et al., 2022; Xue et al., 2024; Mkhari et al., 2023; Mahdavi et al., 2016; Zabot et al., 2016).

Encapsulation efficiency is one of the main properties of encapsulation systems, and can result from the influence of intermolecular interactions between specific chemical groups in the wall material with the active nucleus. The availability of –OH groups in polysaccharides has been proposed to influence the retention of bioactive compounds, explaining, for example, the different behaviors and retention capacities of maltodextrin microparticles related to the DE degrees. Maltodextrins with high DEs contain more –OH groups available for interactions, as evidenced by DE20 maltodextrin microparticles encapsulating 98.5% of anthocyanins (Rosário et al., 2020; Murali et al., 2014). Similarly, short-chain inulin exhibits greater polyphenol and hydrophilic pigment retention due to its higher –OH contents compared to long-chain ones (Gruskiene et al., 2023). Hydrolyzed starch was found to retain higher total polyphenols and anthocyanins, which was attributed to the high number of –OH hydrogen bonds, and dipole-dipole interactions with free –OH in polyphenols and anthocyanin flavylium cations (Rosário et al., 2020). It is worth noting that the increased hydrophilicity of the wall matrix favors the retention of hydrophilic compounds in the amylose helix structure of maltodextrin, as evidenced in microparticles composed of inulin and maltodextrin mixtures displaying greater anthocyanin retention compared to starch alone (Lacerda et al., 2016). The schematic illustration of starch interactions with polyphenols based on the availability of –OH groups is suggested in Figure 7.

The increased DE degree also favors encapsulation governed by hydrophobic interactions since polysaccharides with higher DE have more exposed hydrophobic regions available for binding with hydrophobic bioactive and emulsions (Li et al., 2020; Lee et al., 2017). Corroborating this, hydrolyzed starch was found to retain more than 70% of β-carotenes inside the amylose helix cavity (Rosário et al., 2020), and an illustrative hydrophobic interaction is suggested in Figure 8. Although many studies suggest a strong intermolecular relationship influenced by the number of –OH groups, hydrogen bonding, and hydrophobic interactions between polysaccharides and bioactives, specific analyses, usually carried out by FTIR spectroscopy, confirming these hypotheses are lacking.

On the other hand, in the inulin-alginate microparticles co-encapsulated with polyphenols and betacyanins, FTIR spectroscopy showed hydrogen bonds between the polar groups of the wall materials, such as –OH and –COOH groups of the bioactive compounds (Azevedo & Noreña, 2021). Hydrogen-bonded OH bands have also been evidenced between bioactives from pink pepper (*Schinus terebinthifolius*) extract loaded in maltodextrin, as well as in the alginate-agar microparticles loaded with green tea extract, exhibiting five hydrogen bond-donating OH groups from catechin influencing the hydrogen bonding pattern of intermolecular interaction between wall polysaccharides (Laureanti et al., 2023; Belščak-Cvitanović et al., 2017).

In addition to the polysaccharide-bioactive interaction, bioactive encapsulation can be achieved through the synthesis of microparticles formed by bonds between polysaccharide chains combined in the polymeric wall surrounding the nucleus. Polysaccharide delivery systems based on the mixture of two or more polysaccharide chains have been obtained mainly by ionotropic gelation, in which oppositely charged wall components, such as negatively charged sodium alginate, pectin, and inulin, are linked to the positively charged chitosan as represented in Figure 9. In addition, polysaccharides can be crosslinked to themselves or to reticulant agents such as polyanion TPP, divalent ions $Ca^{2+}$ or $Zn^{2+}$, through electrostatic interactions, leading to the formation of a hydrogel network like an “egg-box” around the bioactive nucleus, as represented by alginate D-galacturonate and D-guluronate chains, forming a network through divalent ions crosslink, schematized in Figure 10 (Lawrie et al., 2007; Wichchukit et al., 2013; Oh et al., 2020; Tan et al., 2024).

A resistant polysaccharide blend microstructure was formed through electrostatic interactions of –COO groups from alginate to protonated $-NH_2$ chitosan groups, capable of encapsulating 89% and 92% of orange peel and black locust flower bioactive extracts, respectively (Savic et al., 2022; Boškov et al., 2024). In addition, blending alginate and pectin resulted in 90% gallic acid encapsulation through physical interactions between the alginate –COOH group and the pectin carbonyl groups (–C=O), according to FTIR spectra, with no signal of linkage between the wall matrix and gallic acid, showing that the bioactive compound can be only surrounded by physical interaction between the polysaccharides (Nájera-Martínez et al., 2023; Pour et al., 2020). In addition, through the formation of a network microstructure by electrostatic interactions, polyphenol and betacyanin were efficiently co-encapsulated at 79% and 89%, respectively, in inulin-alginate microparticles through external ionotropic gelation reaction with $Ca^{2+}$, although interactions between polysaccharide and bioactive were also observed (Azevedo & Noreña, 2021).

It is important to note that polysaccharide-bioactive linkages take place in ionic gelation-based microparticles alongside polysaccharide blends around the nucleus. Hydroxyl groups from alginate, inulin, and gum arabic form intermolecular hydrogen bonds with bioactive core molecules (Patel et al., 2024; Gruskiene et al., 2023; Laureanti et al., 2023). Chitosan can retain active molecules through several intermolecular linkages, including hydrogen bonds between its $-NH_2$ groups and polyphenol –OH groups, electrostatic reactions between opposite-charged chitosan and polyphenol groups, and hydrophobic interactions between hydrophobic chitosan surface areas and the hydrophobic moieties of polyphenols (Charles et al., 2025). On the other hand, molecular interactions between pectin and bioactive molecules are still poorly understood and, thus, referred to as nonspecific ionic interactions. The absence of binding between pectin and polyphenol in microparticles was demonstrated through FTIR spectra analyses, indicating no covalent or other type of bonds with pectin, suggesting that no interactions between pectin and bioactive molecules occurred. Therefore, encapsulating pectin properties appears to be restricted to complexation with other wall polymers, since pectin FTIR spectra do not exhibit new bands or important shifts when compared to those found on empty capsules (Belščak-Cvitanović et al., 2015; Zugic et al., 2025; Vallejo-Castillo et al., 2020).

In sum, several alternatives for the development of microparticles based on edible polysaccharides to entrap polyphenols and pigments have been proposed, showing the ability of those delivery systems for coating hydrophilic and hydrophobic compounds through polysaccharide-polysaccharide and polysaccharide-bioactive physical linkages, with potential to transport, chemical structure, and biological activity at a physiological environment following oral intake.

## 4. MICROENCAPSULATION TECHNIQUES

Several technologies have been developed to load phenolic compounds and pigments into polysaccharide carriers. The encapsulation methods described herein are the most employed for the development of polysaccharide-bioactive delivery systems, the focus of this review. The correct choice of polysaccharide, either single or combined with others, is a crucial step for successful microencapsulation, preservation of loaded compounds under harsh physicochemical conditions, and sustained release. The most employed applied physical methods consist of spray-drying, freeze-drying, and extrusion, while the most employed physicochemical methods, such as coacervation and emulsification, are generally used alongside physical methods (Akpo et al., 2024) (Table 1). However, microencapsulation efficiency depends not only on the applied method, but also on the wall material and the encapsulated core. It is plausible to state that the ideal technique depends on the balance between the characteristics of the active compound, the interaction with the encapsulating polymer, and the functional role intended for the final product. These factors should be considered when choosing the bioactive compound microencapsulation methodology.

### 4.1 Freeze-drying

Freeze-drying, also known as lyophilization, involves dehydration in which a solvent or suspension medium is frozen at reduced temperatures and then undergoes direct sublimation from the solid state to the gaseous state. This process comprises three main steps: freezing, water/solvent sublimation under vacuum in the primary drying phase, and desorption of the small amount of bound water in solids during secondary drying, resulting in a dry material (Boss et al., 2004).

In this process, temperature and pressure are associated with the solid, liquid, and gaseous states of water, in which the intersection of these three phases (known as the triple point) occurs at 0.0098°C and 4.58 mmHg (or 0.00603 atm), making all states occur simultaneously. Freeze-drying takes place below the triple point, eliciting the conversion of ice directly into vapor (Baheti et al., 2010). This process favors the retention of chemical constituents and the stability of bioactive compounds derived from plants, microorganisms, or animal matrices, ensuring a high-quality final product. Regarding structure, lyophilized material exhibits a porous surface and maintains sensory attributes, especially texture (Fan et al., 2019). Because of this, freeze-drying is one of the most employed techniques for the encapsulation of bioactive compounds from edible plants, maintaining their preservation (Ezhilarasi et al., 2013). A beetroot soup rich in polyphenols and dietary nitrate was microencapsulated by freezing drying in starch or maltodextrin as producing spherical particles ranging from 7.94 to 636.34 µm. The microparticles were stable at room temperature for over 90 days, and the sensory beetroot characteristics were preserved, leading to high consumer acceptance. This novel product has been patented according to the Brazilian National Institute of Industrial Property standards under number BR 1020230151965 (Trindade et al., 2023). The freeze-drying methodology has also been employed for the microencapsulation of anthocyanins from juçara fruit (Euterpe edulis) in maltodextrin and arabic gum, retaining 151.68 ± 1.39 mg/100 g and preserving over 80% of polyphenol content (Mazuco et al., 2018).

Freeze-drying, however, presents some disadvantages, such as a higher price compared to conventional drying methods, due to operating costs, as the equipment's cold chamber and vacuum pumps consume high energy. Furthermore, microencapsulation by this method can lead to polyphenol degradation after thawing, as cells disrupted during freezing extrude their hydrolytic enzymes (Shofian et al., 2011; Tan et al., 2015). However, the whole process can release bioactive compounds from food matrices, as described for flavonoids obtained from onions (Pérez-Gregorio et al., 2011). A similar effect was observed in anthocyanin-rich powders obtained by freeze-drying, where anthocyanin retention was 1.5-fold higher than in atomized powders. The freeze-drying effect was compared to spray-drying to encapsulate coffee powder extract in maltodextrin or arabic gum, or a combination of both at the same ratio. Polyphenol and flavonoid retention were optimized following freeze-drying in maltodextrin (Wilkowska et al., 2015).

The antioxidant power of encapsulated compounds can be preserved by freeze-drying (Ozkan et al., 2019). Inulin combined with modified starch, whey protein, and arabic gum has employed to encapsulate geranylgeraniol from annatto (Bixa orellana) oil seed through ultrasound-assisted emulsification followed by freeze-drying. All wall materials combinations retained at least 80% geranylgeraniol antioxidant activity after freeze-drying, where arabic gum-inulin combination preserved 96% of antioxidant power (Silva et al., 2016). In another study, maltodextrin-arabic gum microparticles loaded with sugar maple exhibited an antioxidant capacity up to 241.19±1.21 μM TE/g by the cupric reducing antioxidant capacity (CUPRAC) method throughout storage, preserving over 95% of phenolic and flavonoid compounds after freeze-drying (Yeasmen & Orsat, 2024). These findings reinforce the efficiency of freeze-drying methodology combined with polysaccharides for the stabilization of antioxidant compounds and their bioactivities.

**4.2 Spray drying**

The spray drying methodology transforms liquids into powder by atomization by employing a hot gas injector for drying (Rattes & Oliveira, 2007). The process consists of three main steps: (i) liquid homogenization by an atomizer, (ii) liquid drying using a hot gas carrier to evaporate the solvent, and (iii) collection of the dried particles by cyclones or filters (Schafroth et al., 2012). The liquid is usually presented as a solution, emulsion, or suspension (Gharsallaoui et al., 2007). The injection of the liquid is injected into the drying container carried out through a nozzle or atomizer, producing droplets that evaporate the solvent, (Fatnassi et al., 2014), leading to dry particles which are then collected (Schoubben et al., 2010).

Spray drying is widely applied for the microencapsulation of bioactive food compounds due to its practicality, operation volume, and good cost-benefit ratio (Murugesan & Orsat, 2012). The most common encapsulating agents used in this process include polysaccharides gum arabic, cyclodextrins, maltodextrins, gellan gum, chitosan, whey protein, soy protein, gelatin, and sodium caseinate (Lee & Wong, 2014).

The most significant spray-drying concerns are the high temperatures employed in the process, which can lead to thermal product damage, surface cracks, and loss of volatile compounds (Jafari et al., 2008). Concerning the encapsulation of polyphenol, exposure to oxygen facilitates oxidative reactions, while high temperatures accelerate thermal degradation and promote polymerization, may resulting in loss of functionality (van Golde et al., 2004). In addition, the rapid water outflow for evaporation at high temperatures can result in microparticle surface damage and the premature release of core compounds (Ramírez et al.,

2015).

However, spray drying has been successfully applied for the microencapsulation of various food components, including flavorings, dyes, vitamins, minerals, fats, and oils, and its limitations do not appear to override protective encapsulation effects regarding bioactive compound stability (Pillai et al., 2012). Phenolic extracts from apple peel (Malus domestica) were encapsulated by spray-drying at a maltodextrin and whey protein at the ratio of 8:2 and complemented by the addition of gum arabic at a 6:4 ratio as coating material. Slightly spherical microparticles, from 315 to 719 µm, were obtained at an encapsulation efficiency of up to 83%. The microparticles were used for yogurt enrichment, and a high probiotic activity was noted for up to 15 days of storage, indicating that the encapsulated polyphenols were stable during spray drying and that this process does not reduce the viability of probiotic microorganisms (El-Messery et al., 2019). In another study, an acerola extract (*Malpighia emarginata*) rich in carotenoid and phenolic compounds were encapsulated in gum arabic and maltodextrin by spray drying, generating highly soluble 27 µm spherical microparticles. The microparticles retained 1000 mg GAE/100g and 80 to 99% of anthocyanins from the maple grape skin (Vitis labrusca) in arabic Gum microparticles following the spray drying technique (Kuck & Noreña, 2016). The speed and effectiveness of the method ensure the production of small microparticles with satisfactory morphology, no microbial contamination, and lower costs (Nedovic et al., 2011; Gharsallaoui et al., 2007; Kuck & Noreña, 2016).

### 4.3 Coacervation

Coacervation is widely employed for the microencapsulation of food matrix compounds due to low temperature operating conditions, attenuating thermal degradation and avoiding volatile compounds losses. This process separates liquid-liquid phases in colloidal aqueous solutions, involving one or two oppositely charged polymers. It takes place due to electrostatic interactions, hydrogen bonds, hydrophobic interactions, and polarization effects, or through chemical reactions and crosslinking agents (Xiao et al., 2014). Encapsulation by coacervation can be carried out using a single polymer, *i.e*., gelatin, that undergoes phase separation through changes in temperature, pH, or electrolyte addition, forming a coacervate layer around the core material. Alternatively, it can be employed with two or more oppositely charged biopolymers, typically proteins and polysaccharides, that interact electrostatically to form a coacervate phase, which then encapsulates the active ingredient. The formation of stable microcapsules with suitable properties for load release is influenced by physicochemical conditions, such as pH, ionic strength, and the ratio of polymers used (Napiórkowska & Kurek, 2022). Coacervation allows for high encapsulation efficiency and a controlled release capacity, making it a reliable method for hydrophobic compound encapsulation, conferring versatility for the microencapsulation of several nutraceuticals (Timilsena et al., 2019). One of the disadvantages of coacervation may be related to the instability and degradation of particles during storage, which requires the employment of complementary processes, such as the addition of cryoprotectants, which enhance both storage stability and the dispersibility of coacervates (Muhoza et al., 2023).

Microparticles formed by gelatin and arabic gum, ranging from 35 to 80 µm, allowed for 29 and 38.5% entrapment of anthocyanins from a black raspberry extract. Coacervation microencapsulation resulted in higher stability, reaching up to 23.6% load capacity and preserving product color after two months at 37°C (Shaddel et al., 2018). This methodology has also been used to microencapsulate quercetin in a gelatin-carboxymethylcellulose and curcumin in chitosan microparticles, conferring approximately 90% encapsulation efficiency for quercetin and curcumin, maintaining curcumin antioxidant capacity against the DPPH radical (87%) (Ji et al., 2021; Meiguni et al., 2022).

### 4.4 Extrusion

Microencapsulation by extrusion involves the mixing of encapsulating materials with active compounds in a homogeneous solution, which is subsequently transferred to specialized extrusion equipment through a pointed nozzle with a specific diameter, in which the solution is dripped in a controlled manner. A solution containing the wall material and the active core passes through the nozzle or needle, forming a droplet of controlled shape and size that falls into a hardening solution, usually containing $CaCl_2$, where it then solidifies as microbeads. Calcium chloride enables ionic gelation by interacting with negatively charged encapsulating agents, forming crosslinks that strengthen the microcapsule structure and enhance its stability. The process is fast, allowing for the production of 100-300 microcapsules per second, according to nozzle vibration adjustments. It is worth mentioning that this method has also been reproduced on a small laboratory scale through simple dripping using a syringe and needle. (Silva et al., 2018; Yong et al., 2020; Niño Vasquez et al., 2022). Thus, extrusion is a simple technique to microencapsulate polyphenols into food matrices, forming small and spherical microparticles. However, this technique generates particles with high water content, which can compromise their stability and increase their susceptibility to microbial contamination. This limitation can be solved by drying the microparticles, extending their stability for up to 60 days at room temperature (Silva et al., 2018).

As stated with other microencapsulation techniques, the efficacy of extrusion depends on the choice of polysaccharide coating and adequate processing parameters for the preservation and enhancement of the functional polyphenol properties (Tomé et al., 2022; Haladyn et al., 2021).

The synthesis of hydrogel microparticles combining alginate to different polysaccharides, such as pectin, carrageenan, or chitosan, was found to be very efficient in maintaining the chemical stability of phenolic compounds and preserving the antioxidant capacity of chokeberry (*Aronia melanocarpa*) juice and fruit extracts. Microcapsules prepared with alginate and carrageenan retained 345 mg of phenolic compounds per 100 g of juice and 939 mg per 100 g in the extract-containing capsules. Under these conditions, only 37% and 24% decreases in phenolic content were observed, respectively. When alginate alone was used, about 42% of bioactive compounds were preserved, according to remaining antioxidant activity analyses. Microparticles prepared using alginate and carrageenan showed antioxidant activity superior to 97%, and using alginate and chitosan, reached around 70% of preserved antioxidant activity (Stach & Kolniak-Ostek, 2023).

Anthocyanins obtained from *Hibiscus sabdariffa* L. were encapsulated into pectin employing gelling combined with drip extrusion. The 788 - 1100 μm microparticles retained over 60% of anthocyanins and other antioxidants, even after 35 days of storage at low temperature (De Moura et al., 2018). Extrusion was also employed to encapsulate *Aronia melanocarpa* extracts into hydrogel microparticles prepared using sodium alginate combined with guar gum and chitosan. The microcapsules retained up to 158.08 mg/100 g of polyphenol compounds, preserving 87% polyphenol content after 28 days of storage, with an antioxidant activity of 2.39 mmol TE/100 g, as per an ORAC analysis (Haladyn et al., 2021).

### 4.5 Emulsification

The emulsification method is based on a colloidal system composed of two immiscible hydrophobic and hydrophilic liquid phases stabilized by an interfacial layer composed of an emulsifier that elicits the interaction between phases, forming small droplets dispersed in a non-miscible continuous phase (Flanagan & Singh, 2006). Oil-in-water (O/W) and water-in-oil (W/O) are the most commonly employed emulsions, although multiple emulsion systems, such as double emulsions formed by an internal aqueous phase (W) trapped as small drops within oil droplets (O), which are subsequently dispersed in another aqueous phase (W/O/W) can also be employed (Surh et al.,2006). Droplet emulsions are classified as 'conventional' or macroemulsions when diameters are over 1 μm, while microemulsions range from 10 to 50 μm and droplets or nanoemulsions, between 20 nm and 500 nm (Ramos et al., 2021).

Although emulsion systems designed for microencapsulation and bioactive delivery have demonstrated efficient performance in protecting bioactive compounds and controlling their release to specific sites, conferring stability and bioavailability, recent data have reinforced that emulsification for microencapsulating food bioactive compounds can be enhanced when associated to polysaccharide coatings that improve loaded compound stability and enhance their controlled release, favoring their absorption and potential bioactivity in the human body (Kotta et al., 2012).

Polysaccharides, such as chitosan, carrageenan, starches, and alginate, play a key role in gastrointestinal modulation when included as emulsion-based microparticle components, as they act as structural stabilizers by coating microemulsions while, at the same time, improving digestibility and allowing for the controlled release of bioactive compounds, affecting the absorption and bioavailability of loaded nutrients (Chang & McClements, 2016).

Double-gelled emulsions (W/O/W) formed by a combination of chia mucilage with κ-carrageenan, locust bean gum, and commercial gum acacia plus xanthan blend have been employed for the encapsulation of green tea, demonstrating stable polyphenol compound content for 35 days by maintaining at least 75% of the initial antioxidant activity. The emulsions displayed static stability with no phase separation, and the κ-carrageenan, gum acacia, and xanthan blend formulation demonstrated the best results, promoting the controlled release of green tea during in vitro digestion, where about 80% of phenolic compounds were released under simulated intestinal physicochemical conditions (Guzmán-Díaz et al., 2019).

Similarly, W1/O/W2 double emulsions have been employed to enhance the stability and bioavailability of tea polyphenols using a composite wall material of modified gluten proteins (MEG) and β-cyclodextrin (β-CD) at different ratios, with the best performance observed for MEG: β-CD (2:1) combined with 0.2% xanthan gum or arabic gum. The emulsions were stabilized with 5% polyglycerol polyricinoleate, resulting in an encapsulation efficiency of up to 94%, as well as improved thermal stability and protection under simulated gastric conditions. Furthermore, this system enabled controlled release and achieved a bioavailability of about 60% (Chen et al., 2025).

Chitosan microspheres have been employed to deliver thyme (*Thymus serpyllum L.*) polyphenol compounds by emulsification followed by crosslinking with glutaraldehyde. Microparticles measuring 70-230 μm were obtained using 2% chitosan (w/v) and 0.4% arabic gum (v/v), enabling 67% polyphenol encapsulation efficiency. The microemulsion protected the polyphenols, minimizing their degradation and prolonging their release by 3 h under simulated physicochemical conditions mimicking the gastrointestinal tract. The use of gum arabic provides higher microsphere surface roughness, which may influence their stability and interaction with the gastrointestinal environment (Trifkovic et al., 2014).

Betalains and polyphenols from cactus-pear extract have also been encapsulated in a W/O/W emulsion coated with 60% gum arabic and 40% maltodextrin and used for yogurt enrichment, at 30% of the final mixture. The 20% polysaccharide-based emulsion conferred 22% total phenolic compound retention and 81.8% betalain retention after 36 days of storage. Flavonoids were some of the main identified compounds, although not all phenolics were retained. Furthermore, total polyphenols displayed good stability, and up to 80% of the antioxidant capacity was maintained when exposed to a simulated gastrointestinal environment (Cenobio-Galindo et al., 2019).

## 5. POLYPHENOLS AND PIGMENTS IN POLYSACCHARIDE MICROPARTICLES: CLINICAL PERSPECTIVES

As described here, the development of polysaccharides as bioactive carriers through microencapsulation is a growing technology. The main advantage of the polysaccharide is its biocompatibility, which reduces the risk of adverse reactions, good permeability, facilitates the transport of bioactive molecules, and enables its large-scale use as a carrier due to its low cost. Polysaccharides have been employed in the development of nutraceutical products and have demonstrated relevant potential in the development of medical devices, drug delivery systems, and regenerative materials, promoting advances in biomedicine (Pella et al., 2018).

Polysaccharides are able to transport polyphenols, which exhibit antioxidant and anti-inflammatory properties that contribute to the prevention and treatment of chronic diseases, including cardiovascular and neurodegenerative conditions and certain types of cancer (Potí et al., 2019). These compounds are nutraceuticals permitted for free marketing by health organizations. However, due to limited gastrointestinal stability and absorption, their properties can be improved through polymer encapsulation. In short, when ingested, the polyphenols are hydrolyzed and subjected to chemical modifications such as methylation, alkylation, sulfation, and glucuronidation in the small intestine and liver before entering the circulatory system. Unabsorbed polyphenols, in turn, reach the large intestine and colon, where they can be metabolized by the microbiome into distinct bioactive substances or eliminated in feces and urine (Duda-Chodak et al., 2015). In addition, the interaction of polyphenols with other nutrients, such as proteins, fiber, and minerals, as well as the covalent binding of esterified or conjugated polyphenols to matrix components, can reduce their bioaccessibility and bioavailability, circumventing potential health benefits (Tomas et al., 2017; Kamiloglu et al., 2020; Zhang et al., 2020). These challenges highlight the importance of encapsulation techniques to protect the chemical structure of polyphenols, increasing their biological roles (Kamiloglu et al., 2020). However, the number of *in vitro*, preclinical, and mainly clinical trials is low and should be conducted to evaluate the behavior of microencapsulated polyphenols in the gastrointestinal tract and their biological effects.

In a preclinical assay aimed at improving the stability and bioavailability for oral resveratrol, pectin-alginate microparticles synthesized by ionotropic gelation exhibited resistance to acidic pH, with a slight release of 10% resveratrol in the gastric phase, ensuring its delivery into the small intestine. Microencapsulated resveratrol was tested in mature 3T3-L1 adipocytes to evaluate its influence on triacylglycerol cell content. Both free and encapsulated resveratrol reduced triacylglycerol levels, indicating that stomach-resistant microparticles may represent a promising strategy for oral administration of resveratrol, enabling its use as a dietary supplement and functional food for the treatment of obesity. In addition to ensuring the protection of resveratrol, gastro-resistant encapsulation favors its release into the distal intestinal tract, where absorption is optimized, preserving the bioactivity of resveratrol with regard to the regulation of lipid metabolism (Gartziandia et al., 2018). Resveratrol also exhibited post-digestion antioxidant activity in Caco-2 and HT29-MTX cells, with no cytotoxic effects when loaded onto hydroxymethylcellulose microparticles (Silva et al., 2024).

Preclinical experiments also show increased antiangiogenic effects of caffeic acid loaded in chitosan-β-cyclodextrin microparticles on the reduction of tubular structures formation in human endothelial cells after exposure to hydrogen peroxide ($H_2O_2$)-induced oxidative stress. Microencapsulated caffeic acid looks very promising in the treatment of chronic diseases due to the combination of antioxidant and antiangiogenic activities (Guzmán-Oyarzo et al., 2022).

Chitosan microsponges incorporated into pectin tablets to deliver resveratrol, prevent polyphenol degradation in the upper gastrointestinal tract, maximizing its bioavailability at the target site, showing a remarkable therapeutic effect in the treatment of acid-induced ulcerative colitis. In vivo evidence demonstrated a significant reduction in colonic lesions, necrosis, and inflammation when compared to tissues treated with free resveratrol. In addition, histological analyses demonstrated preservation of mucosal architecture, reduced infiltration of inflammatory cells, and lower ulcer development, indicating that resveratrol loaded in chitosan-pectin microsponges confers cell integrity and colonic environment health. This innovative microencapsulation strategy also demonstrates potential in the treatment of inflammatory bowel diseases, combining controlled-release and increased therapeutic efficacy (Gandhi et al., 2020).

Peanut skin extracts encapsulated in maltodextrin by spray drying significantly reduced postprandial glucose spike in healthy human subjects, as observed in an oral glucose tolerance test. This effect is probably associated with the antioxidant properties of phenolic compounds and their potential interaction with glucose transport and metabolism. Although the exact mechanisms have not yet been elucidated, microencapsulation appears to preserve the bioactivity of the compound and improve its delivery. This strategy may offer promising applications for the dietary management of metabolic conditions such as diabetes while contributing to overall health promotion (Christman et al., 2019).

Studies show that the effects of encapsulation of bioactive compounds go beyond simple physical protection from the gastrointestinal environment, being strongly influenced by the interactions between polymers and active molecules. In pectin and alginate systems, selective release is mainly due to enzymatic degradation, favoring bioavailability in the colon. In hybrid associations that combine cyclodextrins and chitosan, the increase in biological activity is due to the complexation and modulation of the release profile, which maintains more stable concentrations and reduces the loss of the compound. Microspongy structures modulate kinetics in a way that depends on porosity and degradability, expanding the local therapeutic action. Maltodextrin microparticles, on the other hand, act mainly in the modulation of absorption time, without significantly altering bioavailability. Thus, it is understood that the choice of polymer is relevant for physiological performance, as it defines the quality of the interaction with the compound and, consequently, the biological effects obtained.

## 6. Conclusion

Polysaccharides are a promising edible coating for polyphenols and pigments, mainly due to their non-toxicity and low cost. The current scientific literature shows that polysaccharide-polysaccharide and polysaccharide-bioactive compound interactions occur mainly by hydrogen bridges and electrostatic physical interactions, which affect the encapsulation efficiency, the physicochemical and bioactive properties of loaded compounds. Polyphenols and pigments microencapsulated in polysaccharides maintain their chemical and functional characteristics during gastrointestinal route simulations and cell culture models. Furthermore, microencapsulated compounds exhibit good thermostability, light and oxygen resistance, and permeability. Regarding microparticle production, spray drying seems to be more advantageous due to its practicality and cost-effectiveness, while freeze-drying is more suitable for preserving heat-sensitive compounds, although it is a high-energy and time-consuming process.

Chemical techniques like coacervation, emulsification, extrusion, and gelation offer greater control over the particle structure, especially emulsification, which, in combination with polymers such as chitosan and alginate, reduces digestibility and provides the controlled release of bioactive compounds. Investing in hybrid techniques may be an interesting and promising direction, although more experimental support is needed. Most studies focus on encapsulation efficiency, release, and bioaccessibility, but few have addressed issues such as *in vivo* bioavailability and long-term effects. Compiling and reproducing this knowledge and processes on an industrial scale is still challenging, mainly for encapsulation protocols employing multiple coatings or requiring several chemical reaction steps and pH adjustments. In this sense, it seems preferable to explore polysaccharide combinations presenting better solubility in aqueous media, such as modified starches and maltodextrins that do not require the use of chemical agents for solubilization during microparticle synthesis.

## AUTHOR CONTRIBUTIONS

**Liliane Siqueira de Oliveira:** Conceptualization, Data curation, Formal analysis, Investigation, Writing – original draft. **Davi Vieira Teixeira da Silva:** Conceptualization, Data curation, Formal analysis, Investigation, Writing – original draft, Writing – review and editing. **Lucileno Rodrigues da Trindade:** Data curation, Formal analysis, Investigation, Writing – original draft. **Diego dos Santos Baião:** Software, Investigation, Writing – original draft. **Cristine Couto de Almeida:** Data curation, Formal analysis, Investigation, Writing – original draft. **Vitor Francisco Ferreira:** Software, Writing – original draft. **Vania Margaret Flosi Paschoalin:** Supervision, Project administration, Writing – review and editing.

## ACKNOWLEDGMENTS

This study was financed in part by the Fundação de Amparo à Pesquisa do Estado do Rio de Janeiro—Brasil (FAPERJ)—Finance codes of scholarship holders: E-26/200.232/202(Cristine Couto de Almeida); E-26/206.072/2022 (Diego dos Santos Baião); E-26/200.362/2024 (Davi Vieira Teixeira da Silva). Conselho Nacional de Desenvolvimento Científico e Tecnológico—Brasil (CNPq)—Finance code of scholarship holder: 847000/2023-00 (Lucileno Rodrigues da Trindade). Coordenação de Aperfeiçoamento de Pessoal de Nível Superior—Brasil (CAPES)—Finance code of scholarship holder: 8888.957181/2024-00 (Liliane Siqueira de Oliveira).

## Declaration of competing interest

The authors declare no competing of interest.

**Figure captions**

**Figure 1.** Starch sources, their amylose and amylopectin polymers, and maltodextrin obtained from starch hydrolysis.

**Figure 2**. Alginate source and general chemical structure.

**Figure 3.** Pectin sources and general chemical structure.

**Figure 4.** Inulin source and general chemical structure.

**Figure 5.** Chitin source and structure deacetylation forming chitosan.

**Figure 6**. Gum Arabic sources and general chemical structure.

**Figure 7.** Schematic illustration suggesting the interactions between polyphenols and the amylose helix structure of maltodextrin microparticles.

**Figure 8.** Schematic illustration of the possible hydrophobic interaction in the alpha helix cavity using beta-carotene as a model of a hydrophobic bioactive compound encapsulated.

**Figure 9.** Schematic illustration of egg-box structure exemplified by electrostatic interactions between sodium alginate and chitosan around bioactive compounds.

**Figure 10.** Schematic illustration of the egg-box structure around bioactive compounds formed by the linkage between the alginate chains D-galacturonate and D-guluronate with divalent ions.

**Table 1**. Edible polysaccharides for polyphenols and pigments delivery, and the improvement of bioactive physicochemical characteristics

| Encapsulating Polysaccharide | Loaded polyphenols and pigments | Chemical or physical methods | Particle size (μm) | Encapsulation efficiency (%) | Improvements on bioactive compound | Study |
|---|---|---|---|---|---|---|
| Starch | Carotenoids | Gelatinization and freeze-drying | 63 | 75 | - 98% water solubility<br>- Maintenance of antioxidant capacity at 30ºC and under light with a half-life of 37 days | Rosário et al., 2020 |
| | Curcumin/ Resveratrol | ND | ND | 80/ 88 | - 10% release of curcumin in SGD for 2h<br>- 86% release of resveratrol in SID for 10h<br>- 60% of sustained release of antioxidants in SGD and SID for 24 h | Wahab & Janaswamy, 2024 |
| | Anthocyanins | Spray-drying | ND | ND | - 16% release of anthocyanins in SGD and SID for 1h and 12% in intestinal fluid at 2 h | Rosales-Chimal et al., 2023 |
| **Starch + β-cyclodextrin** | Phenolic compounds from pomegranate (*Punica granatum* L.) | Emulsification and spray-drying | 10.16 | 61 | - 49% release of bioactive compounds for 180 min in SID | Cortez-Trejo et al., 2021 |
| **Starch + Inulin + Maltodextrin** | Anthocyanins | Spray-drying | 10 | 67 | - Greater antioxidant power and stability for 38 days under 50°C and light exposure | Lacerda et al., 2016 |
| **Maltodextrin** | Bixin | Emulsification and spray-drying | ND | 86 | - 97% water solubility<br>- 60 days of storage stability at 4º and 28ºC<br>- > 80% of heat stability at 160º C for 15 min | Shridar et al., 2024 |
| | Betalains/ phenolic acid/catechin/ epicatechin/ myricetin | Freeze-drying | ND | 88 | - Higher encapsulation of phenolic acids (44 μg/g) | Mkhari et al., 2023 |
| | Anthocyanins | Spray-drying and freeze-drying | 23.14 | 98 | - 96% water solubility<br>- High antioxidant activity (329 mmol TE/g) | Murali et al., 2014 |
| | Phenolic compounds of stevia extract (*Stevia rebaudiana*) | Emulsification and spray-drying | ND | 87 | - 94% water solubility<br>- Enhanced stability<br>- Stability of phenolic compounds in SGD after 60 min | Zorzenon et al., 2020 |
| **Maltodextrin + Gum Arabic** | Polyphenols | Spray-drying | 27 | 83 | - High retention of total polyphenols and anthocyanins<br>- > 97% water solubility | Kuck et al., 2016 |
| **Inulin** | Hydroxytyrosol and oleuropein from olive leaf extract | Spray-drying | ND | 80 of OH-droxytyrosol<br>79 of oleuropein | - Facilitated colonic delivery (60% release)<br>- Enhanced oleuropein retention after baking (≈77%) and boiling (≈63%)<br>- Improved protection in SID | Duque-Soto et al., 2024 |
| | Betacyanins from beetroot (*Beta vulgaris* L.) extract | Spray-drying | ND | 58 | - Improved color stability in sorbet for 150 days at 18℃<br>- Partial protection of betacyanin in aqueous media<br>- Increased half-life at 4°C | Omae et al., 2017 |
| | Oleuropein from olive leaf (*Olea europaea*) extract | Spray-drying | 10 | 79 | - Retention of oleuropein during baking and boiling<br>- Reduced release in the small intestine (27.5%) indicates delayed bioaccessibility and enhanced oleuropein delivery to the colon<br>- Enhanced stability in intestinal fluid | Pacheco et al., 2018 |
| | Black currant berries (*Ribes nigrum L.*) | Spray-drying | ND | ND | - Increased storage stability of anthocyanins and polyphenols for 12 months at 8℃ and 25℃<br>- Higher antioxidant retention with maltodextrin DE11 | Bakowska-Barczak & Kolodziejczyk, 2010 |
| | Betalains from beetroot (*Beta vulgaris* L.) extract | Freeze-drying | ND | ND | - Improved water solubility<br>- Reduced hygroscopicity<br>- Promote thermal stabilization<br>- Preservation of antioxidant activity and polyphenols during storage | Flores-Mancha et al., 2020 |
| | Anthocyanins from chokeberry (*Aronia berries*) extract | Spray-drying | 17 | 88 | - 90% water solubility<br>- Reduced anthocyanin degradation during storage<br>- Improved stability under light and oxygen | Pieczykolan & Kurek, 2019 |
| **Inulin or Acetylated Inulin** | Gallic acid | Spray-drying | ND | 83 inulin<br>75 acetylated inulin | - Acetylated inulin reduced burst release effect and slowed gallic acid release, non-Fickian (anomalous) diffusion behavior | Robert et al., 2012 |
| **Inulin + Modified Starch** | Oregano (*Origanum vulgare L.*) extract | Emulsification and spray-drying | 151 | 66 | - Thymol retention (84%)<br>- High entrapment efficiency (92%)<br>- Improved thermal stability up to 220°C | Zabot et al., 2016 |
| **Inulin + Maltodextrin** | Anthocyanin from mangosteen (*Garcinia mangostana*) | Spray-drying | 13 | 81 | - High thermal stability (<200°C mass loss)<br>- 80% anthocyanin retention in 120 min at 70°C<br>- Increased stability and shelf-life when incorporated into yoghurt | Sakulnarmraet al., 2022 |
| **Inulin + Sodium Alginate** | Betacyanins (*Bougainvillea glabra*) bracts | External ionic gelation and extrusion | 4.3 mm (Feret's diameter) | 89 | - Increasing inulin content from 10% to 20% enhanced the rheological properties of dispersions<br>- Spherical shape<br>- Increased mechanical resistance<br>- Thermal stability up to 200°C and antioxidant protection | Azevedo & Noreña, 2021 |
| **Inulin or Gum Arabic + Maltodextrin** | β-carotene extracted from mango peel (*Mangifera indica*) | Spray-drying | ND | 87 | - Enhanced β-carotene stability (>68% recovery after SGI)<br>- 2-fold increase in bioaccessibility with inulin addition<br>- Improved cellular uptake of β-carotene in Caco-2 cells | Cabezas-Terán et al., 2023 |
| **Inulin + Maltodextrin + WPI** | Betalains from beetroot (*Beta vulgaris* L.) extract | Spray-drying | ND | 88 - 95 | - Increased antioxidant activity<br>- Enhanced thermal stability (~200°C) | Carmo et al., 2017 |
| **Inulin + Maltodextrin + Gum Arabic** | Anthocyanins from Roselle (*Hibiscus sabdariffa L.*) | Spray-drying and freeze-drying | ND | 92 by spray-dried, 95 by freeze-dried | - Higher retention of anthocyanins and antioxidant activities with MD/GA<br>- > 94% solubility in all delivery systems | Nguyen et al., 2022 |
| **Inulin + Maltodextrin + Gum Arabic** | Lycopene from tomato (*Lycopersicon esculentum*) extract | Spray-drying | ND | ND | - Improved lycopene stability under storage<br>- Enhanced intestinal delivery<br>>50% bioaccessibility when incorporated into yogurt | Corrêa-Filho et al., 2022 |
| **Alginate** | Catechins | Ionic gelation, emulsification, and extrusion | 131 | 42 | - 63% sustained catechins release after 2 h in a pH 7 solution | Kim et al., 2016 |
| **Alginate + Carrageenan** | Phenolic compounds | Extrusion | ND | 97 | - Retention of 939 mg of phenolic compounds per 100 g in microcapsules containing phenolic extract<br>- High antioxidant capacities after one month of storage at 4°C | Stach & Kolniak-Ostek, 2023 |
| **Alginate + Pectin** | Resveratrol | Ionic gelation by atomization in a peristaltic pump | 1.4 | 41 | - > 90% of stability for 2 h in pH 1.2 solution;<br>- 70% of sustained release for 24 h in tri-sodium phosphate at pH 7.4 | Gartziandia et al., 2018 |
| | Black locust flower extract | Ionic gelation and extrusion | 228 | 92.5 | - 60% sustained antioxidants release for 24 h in gastric and intestinal fluids<br>- Increased antioxidant activity in SGD and SID | Boškov et al., 2024 |
| | Phenolic acids from orange peel | Ionic interaction, ionic gelation, and extrusion | 252 | 89 | - 82% of sustained antioxidants release for 24 h and increased antioxidant activity in SGD and SID | Savic et al., 2022 |
| **Alginate + Chitosan** | Quercetin | Electrostatic complexation and extrusion | ND | 86 | - Intact microparticle structure after 8h in a pH 1.2 solution<br>- Sustained release for 24 h in a pH 7.4 solution | Frent et al., 2022 |
| | Quercetin | Emulsion-assisted external ionic gelation and freeze-drying | 25 - 79 | 53 (with the combination of inulin and chitosan) | - Retention of 80% quercetin after SID<br>- Delayed release during colonic fermentation<br>- Increased structural integrity of microspheres | Liu et al., 2022 |
| **Chitosan** | Polyphenols from *Thymus serpyllum L.* extract | Emulsion and crosslinking oven processing at 50 °C | 68 - 230 | ND | - pH-responsive prolonged release<br>- Higher release in SGD and SID vs water<br>- Swelling-controlled release<br>- Maintenance of antioxidant activity<br>Increased matrix crystallinity | Trifkovic et al., 2015 |
| | Anthocyanin from blueberry (*Vaccinium caesariense*) extract | Ionic gelation and spray-drying | 0.06 | 94 | - High anthocyanin recovery<br>- Improved pH stability (especially at pH 7.4)<br>- Reduced release at neutral pH | Wang et al., 2016 |
| **Chitosan + Fish Gelatin** | Phenolic compounds from apple pomace (*Malus domestica*) extract | Spray-drying | 203 | 88.6 | - Preserved ≈70% antioxidant activity after 5 weeks at 35°C<br>- Improved stability and dispersion<br>- Potential for Pickering emulsions | Moradi et al., 2024 |
| **Chitosan + Inulin + WPI** | Anthocyanins from black currant (*Ribes nigrum*) and Lactic acid bacteria (*Lactobacillus casei* 431®) | Spray-drying | < 5 (small spherosomes)<br><br>≈20 - 337 (large coaservates) | 95.46 | - Stability in gastric fluid and sustained release in intestinal fluid<br>- Stability over 90 days at 4°C<br>- Antioxidant and antidiabetic activity preservation | Enache et al., 2020 |
| **Gum Arabic** | *Ruellia tuberosa* L. and *Tithonia* diversifolia leaves powder | Spray-drying | 145 | 84 | - Alpha-amylase inhibition ($IC_{50}$: 54 µg/mL)<br>- Increased antioxidant activity (DPPH) | Almayda et al., 2024 |
| **Gum Arabic + WP** | Flavonoids from ponkan peel (*Citrus reticulata Blanco*) | Arabic gum emulsion in soybean oil and WP in water Spray-drying | 1.42 | 97 | - $ABTS^{\cdot +}$ and DPPH inhibition | Hu et al., 2017 |
| **Gum Arabic + Coconut (*Cocos Nucifera L.*) + milk Whey** | Curcumin | Spray-drying | < 22 | 94.5 | - Over 88% water solubility<br>- 72 days of storage stability at 50°C (15%) | Adsare & Annapure, 2021 |
| **Gum Arabic +** | Anthocyanins from barberry (*Berberis vulgaris*) extract | Spray-drying | ND | 96 | - 91% water solubility | Mahdavi et al., 2016 |
| **Maltodextrin** | Flavonoids | Freeze-drying | 7.2 | 95 | - Retention of phenolic and flavonoid for one month of storage under refrigeration and room temperature<br>- > 89% water solubility | Yeasmen & Orsat, 2024 |
| **Gum Arabic + Inulin** | Buriti (*Mauritia flexuosa*) oil | Freeze-drying | 7.5 | 93 | - 92% β-carotene retention<br>- 72% water solubility | de Oliveira et al., 2022 |
| | Geranylgeraniol | Freeze-drying | 1.1 | 96 | - 96% of antioxidant activity stability maintenance (ORAC and DPPH•) | Silva et al., 2016 |
| **Gum Arabic + Gelatin** | Anthocyanins | Freeze-drying | 63 | ND | - 48% - 70% of anthocyanins retention after 60 days at 7°C | Shaddel et al., 2017 |
| **Gum Arabic + Chitosan** | Polyphenols | Emulsification | ND | 67 | - Reduced polyphenols degradation and 3 h prolonged release under simulated digestion | Trifkovic et al., 2014 |
| **Pectin** | Resveratrol | Ionic interaction and cross-linking of pectin-glutaraldehyde by extrusion | 902 | 98 | - 93% of stability for 5h in SGD and SID<br>- > 90% of storage stability at 4˚, 25˚ and 40˚C for180 days<br>- 5h increment to reach the plasma of Sprague–Dawley rats | Das et al., 2011 |
| **Pectin + Alginate** | Gallic acid | Ionic gelation and air-drying | 1.3 | 89 | - 15 - 35% of sustained cumulative release in pH 4 of HCl solution | Nájera-Martinez et al., 2023 |

ND – not determined, $CaCl_2$ – calcium chloride, $D_h$ – hydrodynamic diameter, DPPH – 2,2-diphenyl-1-picrylhydrazyl, HCl – hydrochloric acid, $IC_{50}$ – half-maximal inhibitory concentration, $Zn^{2+}$ – zinc ion, LC – loading capacity, SGD – simulated gastric digestion, SID – simulated intestinal digestion, TE – Trolox equivalents, WP – whey protein, WPI – whey protein isolate